\documentclass[twoside,twocolumn,9pt]{article}
\usepackage{extsizes}
\usepackage[super,sort&compress,comma]{natbib}
\usepackage[version=3]{mhchem}
\usepackage[left=1.5cm, right=1.5cm, top=1.785cm, bottom=2.0cm]{geometry}
\usepackage{balance}
\usepackage{mathptmx}
\usepackage{sectsty}
\usepackage{graphicx}
\usepackage{lastpage}
\usepackage[format=plain,justification=justified,singlelinecheck=false,font={stretch=1.125,small,sf},labelfont=bf,labelsep=space]{caption}
\usepackage{float}
\usepackage{fancyhdr}
\usepackage{fnpos}
\usepackage[english]{babel}
\addto{\captionsenglish}{%
  
}
\usepackage{array}
\usepackage{droidsans}
\usepackage{charter}
\usepackage[T1]{fontenc}
\usepackage[usenames,dvipsnames]{xcolor}
\usepackage{setspace}
\usepackage[compact]{titlesec}
\usepackage{hyperref}

\usepackage{epstopdf}

\definecolor{cream}{RGB}{222,217,201}

\definecolor{purple}{rgb}{0.7,0,0.7}

\definecolor{lightblue}{rgb}{0.6,0.6,1}
 \definecolor{dgreen}{rgb}{0.1,0.5,0.0}

\newcommand{\rs}[1]{{\color{dgreen} #1}}

\begin{document}

\pagestyle{fancy}
\thispagestyle{plain}
\fancypagestyle{plain}{
\renewcommand{\headrulewidth}{0pt}
}

\makeFNbottom
\makeatletter
\renewcommand\LARGE{\@setfontsize\LARGE{15pt}{17}}
\renewcommand\Large{\@setfontsize\Large{12pt}{14}}
\renewcommand\large{\@setfontsize\large{10pt}{12}}
\renewcommand\footnotesize{\@setfontsize\footnotesize{7pt}{10}}
\makeatother

\renewcommand{\thefootnote}{\fnsymbol{footnote}}
\renewcommand\footnoterule{\vspace*{1pt}%
\color{cream}\hrule width 3.5in height 0.4pt \color{black}\vspace*{5pt}}
\setcounter{secnumdepth}{5}

\makeatletter
\renewcommand\@biblabel[1]{#1}
\renewcommand\@makefntext[1]%
{\noindent\makebox[0pt][r]{\@thefnmark\,}#1}
\makeatother
\renewcommand{\figurename}{\small{Fig.}~}
\sectionfont{\sffamily\Large}
\subsectionfont{\normalsize}
\subsubsectionfont{\bf}
\setstretch{1.125} 
\setlength{\skip\footins}{0.8cm}
\setlength{\footnotesep}{0.25cm}
\setlength{\jot}{10pt}
\titlespacing*{\section}{0pt}{4pt}{4pt}
\titlespacing*{\subsection}{0pt}{15pt}{1pt}

\fancyfoot{}
\fancyfoot[RO]{\footnotesize{\sffamily{1--\pageref{LastPage} ~\textbar  \hspace{2pt}\thepage}}}
\fancyfoot[LE]{\footnotesize{\sffamily{\thepage~\textbar\hspace{3.45cm} 1--\pageref{LastPage}}}}
\fancyhead{}
\renewcommand{\headrulewidth}{0pt}
\renewcommand{\footrulewidth}{0pt}
\setlength{\arrayrulewidth}{1pt}
\setlength{\columnsep}{6.5mm}
\setlength\bibsep{1pt}

\makeatletter
\newlength{\figrulesep}
\setlength{\figrulesep}{0.5\textfloatsep}

\newcommand{\topfigrule}{\vspace*{-1pt}%
\noindent{\color{cream}\rule[-\figrulesep]{\columnwidth}{1.5pt}} }

\newcommand{\botfigrule}{\vspace*{-2pt}%
\noindent{\color{cream}\rule[\figrulesep]{\columnwidth}{1.5pt}} }

\newcommand{\dblfigrule}{\vspace*{-1pt}%
\noindent{\color{cream}\rule[-\figrulesep]{\textwidth}{1.5pt}} }

\makeatother

\twocolumn[
  \begin{@twocolumnfalse}
\vspace{1em}
\sffamily
\begin{tabular}{m{0cm} p{17.5cm} }

& \noindent \LARGE{\textbf{Silo discharge of mixtures of soft and rigid grains}} \\
\vspace{0.3cm} & \vspace{0.3cm} \\

 & \noindent\large{Jing Wang$^{a*}$, Bo Fan$^{b*}$, Tivadar Pong\'o$^{c*}$, Kirsten Harth$^a$, Torsten Trittel$^a$, Ralf Stannarius$^{a\dag}$,
Maja Illig,
Tam\'as B\"orzs\"onyi$^b$,  Ra\'ul Cruz Hidalgo$^c$} \\
\vspace{0.3cm} & \vspace{0.3cm} \\

& \noindent \normalsize{
We study the outflow dynamics and clogging phenomena of mixtures of soft, elastic low-friction spherical grains and  hard frictional spheres of similar size in a quasi-two-dimensional (2D) silo with narrow orifice at the bottom. Previous work has demonstrated the crucial influence of elasticity and friction on silo discharge. We show that the addition of small amounts, even as low as 5\%, of hard grains to an ensemble of soft, low-friction grains
already has significant consequences. The mixtures allow a direct comparison of the probabilities of the different types of particles to clog the orifice. We analyze these probabilities for the hard, frictional and the soft, slippery grains on the basis of their participation in the blocking arches, and compare outflow velocities and durations of non-permanent clogs for different compositions of the mixtures.
Experimental results are compared with numerical simulations. The latter strongly suggest a significant influence of the inter-species particle friction.} \\

\end{tabular}

 \end{@twocolumnfalse} \vspace{0.6cm}
  ]

\renewcommand*\rmdefault{bch}\normalfont\upshape
\rmfamily
\section*{}
\vspace{-1cm}


\footnotetext{\textit{$^{a}$~Institute of Physics, Otto von Guericke University, Department of Nonlinear Phenomena,
D-39106 Magdeburg, Universit\"ats\-platz 2, Germany.$^b$ Institute for Solid State Physics and Optics, Wigner Research Center for Physics,
P.O. Box 49, H-1525 Budapest, Hungary,
$^c$ F\'isica y Matem\'atica Aplicada, Facultad de Ciencias, Universidad de Navarra, Pamplona, Spain, $^\ast$ shared first authorship}}

\footnotetext{$\dag$~ralf.stannarius@ovgu.de}

\section{Introduction}
Storage of granular materials in silos and hoppers has an evident advantage over other containers for the processing of these materials in agriculture,
chemical industry, construction industry and many other branches: Material is stowed into the container through a top orifice, yet it can be withdrawn from the storage device through an orifice at the bottom simply using gravity forces. No additional mechanical device is needed to maintain the outflow. However, one of the problems with these storage bins is congestion of the orifice, so-called clogging, which can occur even if the orifice diameter is much larger than the largest spatial extension of the individual grains. Particles can form stable arches (in 2D) or domes (in 3D) above the outlet and block further outflow. Intervention from outside is required to re-trigger discharge.
From a physical point of view, this process is insufficiently understood even today, despite of hoppers being in use for millenia in human history.
The flow of grains, even in the simplest form of hard monodisperse spheres, has retained many mysteries.
Even the simplest problem of the outflow of monodisperse ensembles of spheres
is still an active field of research.
%
Through large enough orifices, such grains flow at
constant rates given by geometrical and physical parameters. Flow rates have been derived from theoretical models
\cite{Franklin1955,Beverloo1961,Neddermann1982,Mankoc2007} and the predictions were tested in numerous experiments (e.g.  \cite{Mankoc2007,Aguirre2010, Sheldon2010,Janda2012,Koivisto2017,Wilson2014}). When the outlet diameter is
small (less than about 5 particle diameters) \cite{To2005,Zuriguel2005,Thomas2015}, hard spheres tend to form clogs after some time at the orifice. These block further
outflow until they are destroyed by shaking the container, by applying air flushes or by other mechanical disturbances.
Spontaneous arch formation~\cite{Tang2011,Zuriguel2014_1}, the preceding kinetics
\cite{Rubio-largo2015}, as well as the inherent force distributions~\cite{Hidalgo2013,Vivanco2012} have been analyzed in the literature.
Identical hard spheres are an idealized special system that has been considered in most of the experimental and theoretical studies.
Since there are no principal differences between clogging of 2D and 3D silos, the former are often preferred in experiments because they offer the study of the inner dynamics and structure formation in the container with non-invasive observation methods.

Recently, it was found that soft particles with low friction differ
considerably from common hard, frictional grains in their static and dynamic behavior in silos
\cite{Hong2017,Ashour2017b}. The critical ratio $\rho$ of orifice size and
particle size below which clogging sets in is much smaller than for
rigid grains. In addition, non-permanent, intermittent clogs are
observed which spontaneously dissolve after some time. Such features
are otherwise observed only in vibrated silos~\cite{Mankoc2009,Guerrero2018}, in colloidal systems \cite{Marin2018,Cruz2018,Souzy2020} or in active
matter \cite{Zuriguel2014_1}. In the earlier experiments with
soft, slippery hydrogel grains, it remained unexplained whether the
softness of the particles (elastic modulus of the order of a few
dozen kPa) or the low friction coefficient (in the order of 0.03),
or a combination of both causes the quantitative differences to
hard, frictional grains.
Previous work on pure monodisperse hydrogel sphere (HGS) ensembles \cite{Harth2020} has identified the viscoelastic properties of these spheres as one important feature that causes qualitatively new features of the discharge through narrow orifices.

In practice, homogeneous granular ensembles are often the exception, even in industrial processes. One usually deals with materials that are non-uniform in size, shape, surface structure, or other properties.
This motivated us to study mixtures of particles that differ in their elastic and frictional properties but are otherwise very similar. It turns out that an addition of even small portions of rigid particles to HGS ensembles has dramatic influence
on silo discharge behavior.

This paper describes an experimental and numerical study of the effects of doping soft HGS ensembles with hard, frictional particles of the same size and weight. We will add up to 10 \% of the latter to the pure hydrogels and focus on three aspects:\\
(1) What is the influence of doping on the silo discharge characteristics?\\ (2) How is the concentration $x_{\rm hf}$ of hard grains in the mixture represented in the composition of the blocking arches? This will allow us to extract the probabilities that hard or soft grains complete a blocking arch and cause clogging.\\ (3) Does the composition of the mixtures affect the  outflow rates?

A quasi-two-dimensional setup with one layer of beads between two vertical glass plates is used. The mass of the discharged material is recorded by means of a balance beneath the orifice. Particle arrangements and flow inside the container are monitored by video imaging. In addition, numerical simulations are performed and compared to the experiment.

\section{Experimental setup and materials}
\label{sec:Setup}

\begin{figure}[htbp]
    \centering
    \includegraphics[width=0.75\columnwidth]{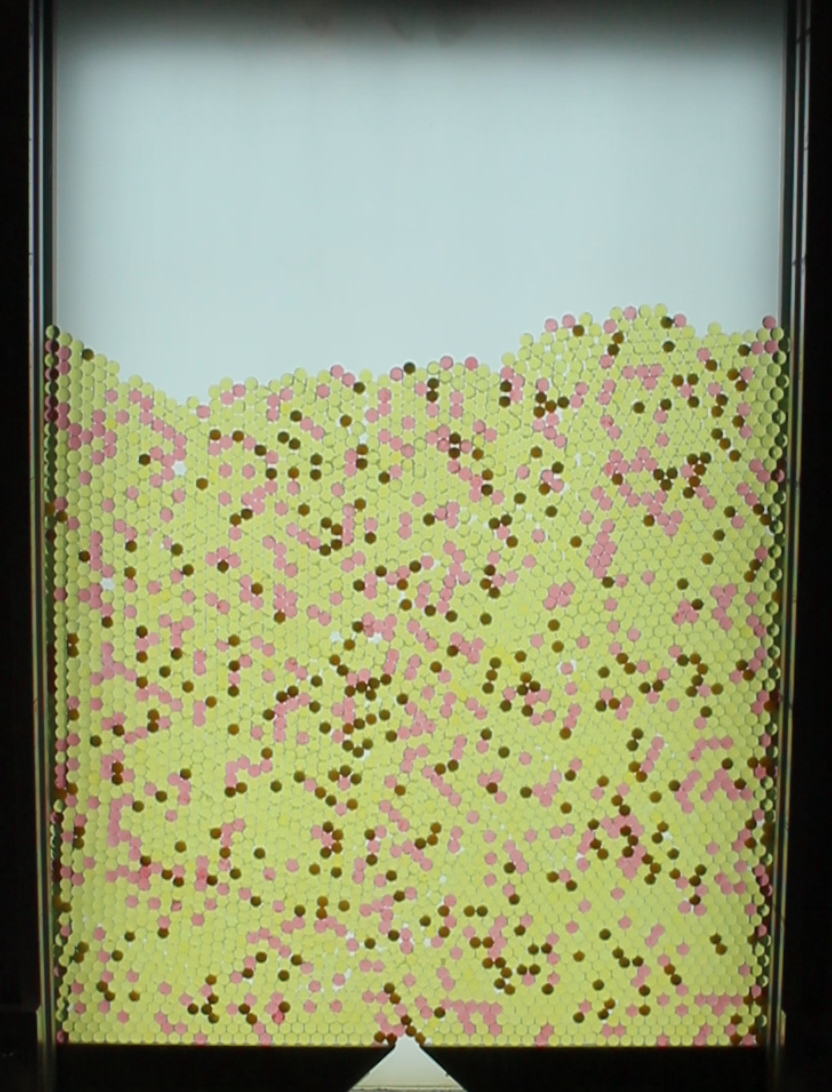}\vspace{3mm}
    \includegraphics[width=0.75\columnwidth]{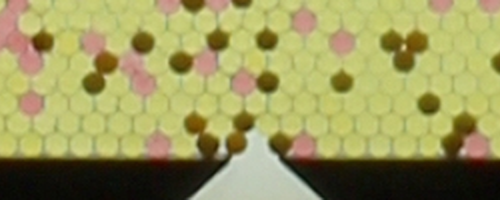}
    \caption{Top: Image of the container filled with a mixture with 10~\% hard frictional beads. Yellow and pink colored hydrogel spheres are equivalent, airsoft bullets appear dark. Bottom: Zoom into the region near the orifice at the bottom, in a clogged state. The orifice  in these images is $D=11$~mm, slightly less than two particle diameters.}
    \label{fig:setup}
\end{figure}

The setup consists of a flat container of 80 cm height, 40 cm
and slightly more than 6 mm depth. In
the images shown in Fig.~\ref{fig:setup}, vertical aluminum rails that support the front and rear glass plates hide 3 cm of the container
interior on the left and right sides. Only 34 cm are visible. At
the bottom, a rectangular opening of variable width can be adjusted
with two horizontal sliders.

The container is filled with a mixture of soft, low-friction hydrogel spheres and hard frictional (HF) plastic airsoft ammunition. The concentration of HF plastic grains is low, typically 5$\%$ or 10$\%$. We define the aspect ratio $\rho$ as the quotient of the orifice  $D$ and the particle diameter $d$. In the present study, this ratio is in the range $1.7 < \rho < 2.2$. We note that at such small orifice sizes a pure sample of hard frictional grains would almost immediately clog. At large enough orifice sizes ($\rho >3$), the behaviour of our mixtures is practically identical to the pure hydrogel samples.

Both species have densities of approximately 1020 kg/m$^3$. The 6 mm diameter airsoft beads were obtained from commercial providers. They are made from plastic, are perfectly monodisperse and they are hollow. The latter
feature is irrelevant here. The friction coefficient of the airsoft beads is
0.3. They are incompressible and can be considered rigid. The hydrogel spheres
were obtained in dried state from a commercial provider ({\em
Happy Store, Nanjing}). They were swelled in salted water with a
NaCl concentration that determined the final radius of approximately
6.5 mm, with a
polydispersity of about 3 \%. The mass of each hydrogel sphere is about 0.15~g. They have a friction coefficient one order of magnitude
lower than the HF spheres, but the elastic modulus is of the order of only 50~kPa to
100~kPa. These particles are incompressible as well, but they deform
slightly in the silo under the weight of the overlying grains (see
Fig.~\ref{fig:setup}, bottom). A rough estimate is that the Hertzian contacts between hydrogel spheres at the bottom of the container indent the grains by approximately 1.3 mm under the weight of the full silo.

 The 2-dimensional (40 x 80 cm) cell can accommodate about 10,000 grains corresponding to a weight of $\approx 1.5$ kg. The cell is extended at the top with an additional 3-dimensional container, which can hold  extra granular material.

\section{Avalanches and clogging probabilities}

\begin{figure}[htbp]
    \centering
    \includegraphics[width=0.48\textwidth]{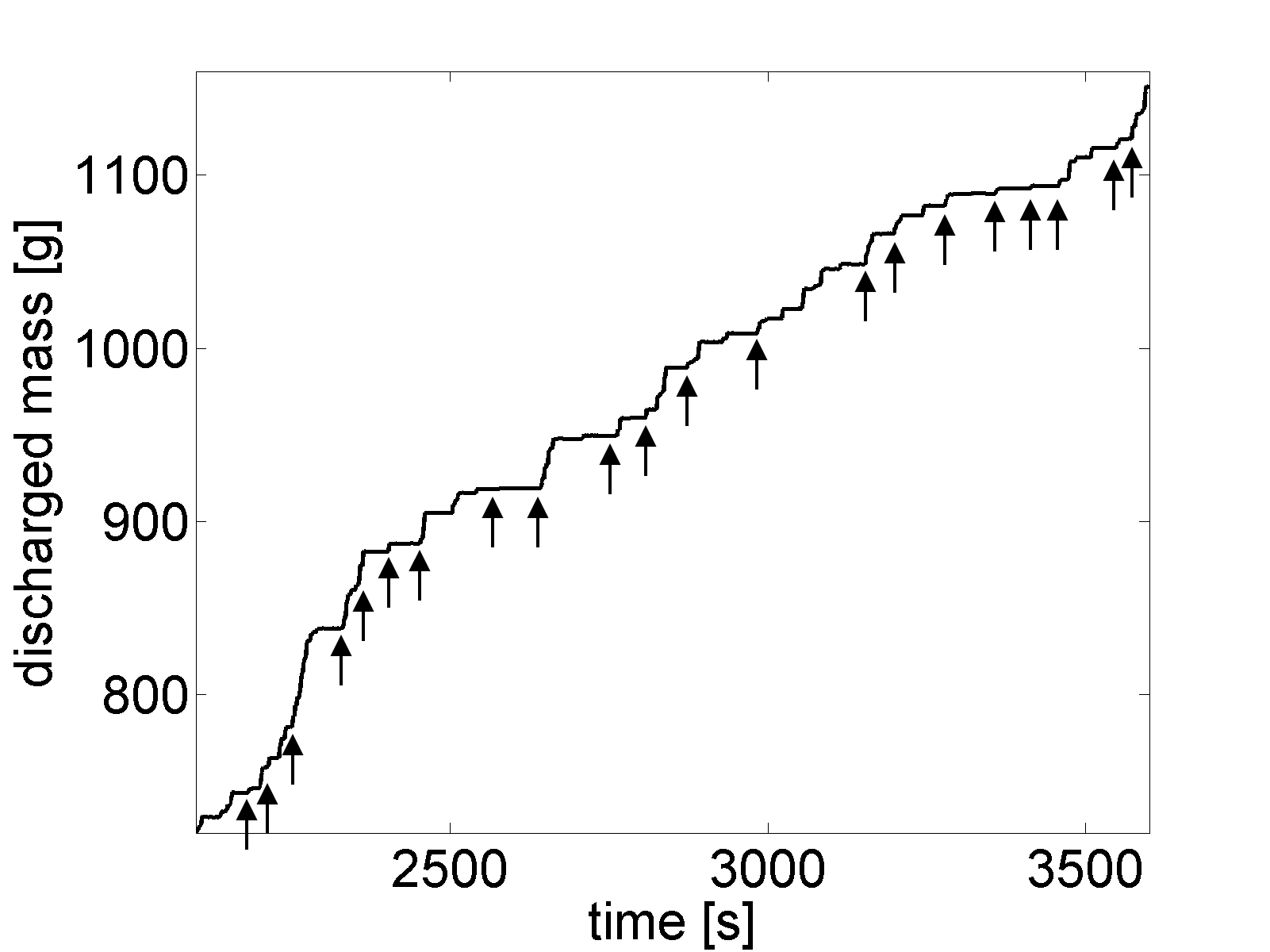}
    \caption{Part of a measured time dependence $m(t)$ of the mass of granular material discharged from a quasi-2D silo. The mixture contained 10 \% of rigid air soft balls with 90 \% hydrogel spheres, both with 6 mm diameter. The orifice size was 11 mm. Some clogs (plateaus) dissolved spontaneously, clogs labeled by arrows were destroyed by an air flush through the orifice.}
    \label{fig:mt}
\end{figure}

Avalanche sizes are a key figure of merit for silo discharge.
Figure \ref{fig:mt} shows an example of the discharged mass curve for a mixture of hard and soft spheres, filled into a silo with a narrow orifice.
Plateaus in this curve either represent non-permanent interruptions of the flow that dissolve spontaneously, or clogs that were destroyed by air flushes (arrows).

The mean number of grains discharged during an avalanche is directly related to the probability that a particle completes a blocking structure at the orifice. In the geometry we use in our experiments, with $\rho$ of the order of 2, these clogs are formed by very few grains, four on average. Thus, the experiment provides
favorable conditions to observe the microscopic dynamics (on the particle level)  at the opening, and on the other hand, the choice of the soft hydrogel material guarantees the formation of comparably large avalanches for a reasonable statistics at
these small aspect ratios.
The relation between the probability of particles completing a clog at the outlet and the size of avalanches is given in Appendix A.

In the experimental determination of avalanche sizes, one encounters a problem which is related to a peculiar feature of the hydrogels identified already in the pure system \cite{Harth2020}: The soft, viscoelastic material has the tendency to form interruptions of the outflow that can dissolve spontaneously after some time, particularly for small orifice sizes. In Figure \ref{fig:mt},
some plateaus indicate non-permanent interruptions of the flow that dissolve spontaneously.
There is no clear criterion to discriminate the end of an avalanche and the beginning of a non-permanent clog from mere fluctuations of the outflow rate.  If one analyzes the time delays between subsequent grains passing the orifice, there
is no obvious threshold that may serve to identify and mark non-permanent clogs, and distinguish them from fluctuating outflow. Short interruptions of the outflow are continuously distributed. Technically, it is therefore justified to regard an outflow process as one single avalanche, unless a permanent clog is reached. In Fig.~\ref{fig:mt}, the nearly 3 minute period
after the air flush at $t=2980$ s represents an example of an avalanche that has intermittent non-permanent breaks.

For practical reasons, one may nevertheless be interested in the distribution of phases where the material is flowing and phases where the outflow is interrupted. This is the standard procedure for the description of living or externally agitated systems (e.~g.~\cite{Zuriguel2014}). Thus, we introduce an {\em ad hoc} criterion, the interruption of the outflow lasting one second or longer, to separate avalanches. However, it was shown earlier that the arbitrary selection of a threshold may influence the statistics considerably \cite{Cruz2018}. Since we treat all mixtures with the same model, our arbitrary choice of the threshold may be still justified.

\section{Experimental results}

\subsection{Pressure characteristics}
\label{sec:Exp:pressure}

\begin{figure}[htbp]
    \centering
    \includegraphics[width=0.9\columnwidth]{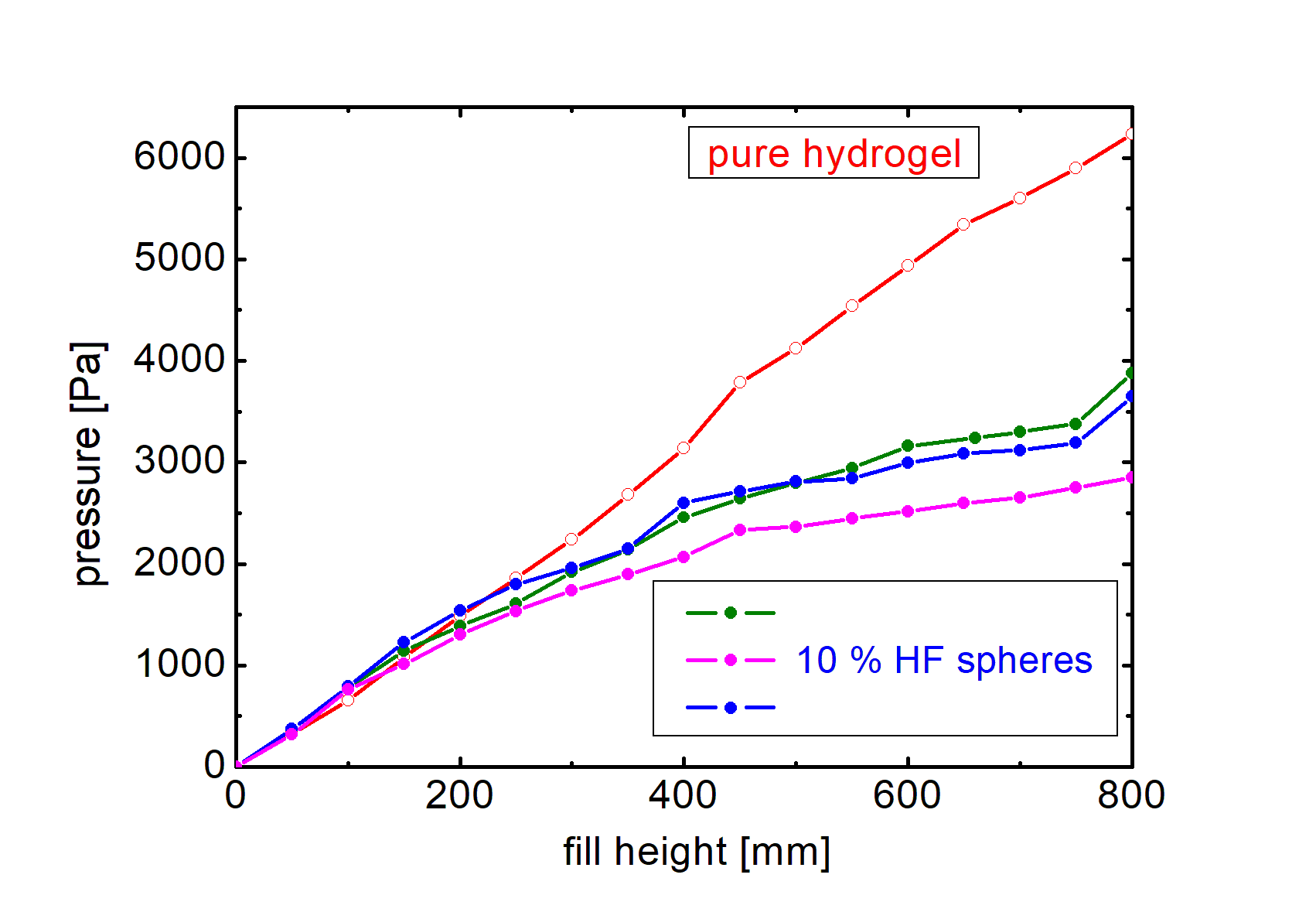}
    \caption{Pressure at the bottom of the container at the position of the outlet for pure hydrogel system (open circles), and three independent measurements of a  mixture with 10\% HF spheres (filled circles) (cf. numerical simulation data in Fig.~\ref{fig:sim_force_vs_height}).
    Lines guide the eye.}
    \label{fig:pressure}
\end{figure}

It was shown earlier that the low-frictional hydrogel shows an almost hydrostatic characteristics $P(h)$ of the pressure
$P$ at the bottom of a quasi-2D container filled up to a height $h$ \cite{Ashour2017b}.
In contrast, the hard, frictional grains exhibit the typical saturation of the pressure
\cite{Janssen1895} at a fill height of several cm \cite{Ashour2017b}. Figure \ref{fig:pressure} shows that the pressure
characteristics of the pure hydrogel sample is changed
significantly by addition of a small amount $x_{\rm hf}$ (10 \%) of HF spheres. The pressure in the mixtures clearly deviates from hydrostatic behavior. The data were obtained by measuring the force on a short (4 cm) horizontal bar that replaced part of the bottom container border.
The weight of the material in upper layers is at least partially transferred to the container walls. The pressure characteristics differ slightly in individual runs, but the general trend is seen in all three graphs.
This continuous pressure increase with fill height has direct consequences for the discharge characteristics of the mixtures, as will be demonstrated below.

One consequence of the increasing pressure towards the bottom of the silo in combination with
the low elastic modulus of the hydrogels is that the packing fraction $\phi$ increases towards the bottom.
The closest packing of spheres with diameter $d$ in a quasi-2D hexagonal lattice within a layer of thickness $d$ is $\phi_{\max} = \pi/\sqrt{27}\approx 0.604$. This is indeed the mean packing fraction of pure hydrogel spheres, since they form a nearly defect-free hexagonal lattice in the depth of the granular bed. At the bottom, they even reach packing fractions up to about 0.65 where they are squeezed out of their original sphere shape. In the very top layers, owing to imperfections, the packing fraction drops to about 0.5. This may have some consequences for the outflow rates discussed in the next section.

\subsection{Flow rate and clog duration}
\label{sec:flowrate}
\begin{figure}[htbp]
    \centering
    \includegraphics[width=0.99\columnwidth]{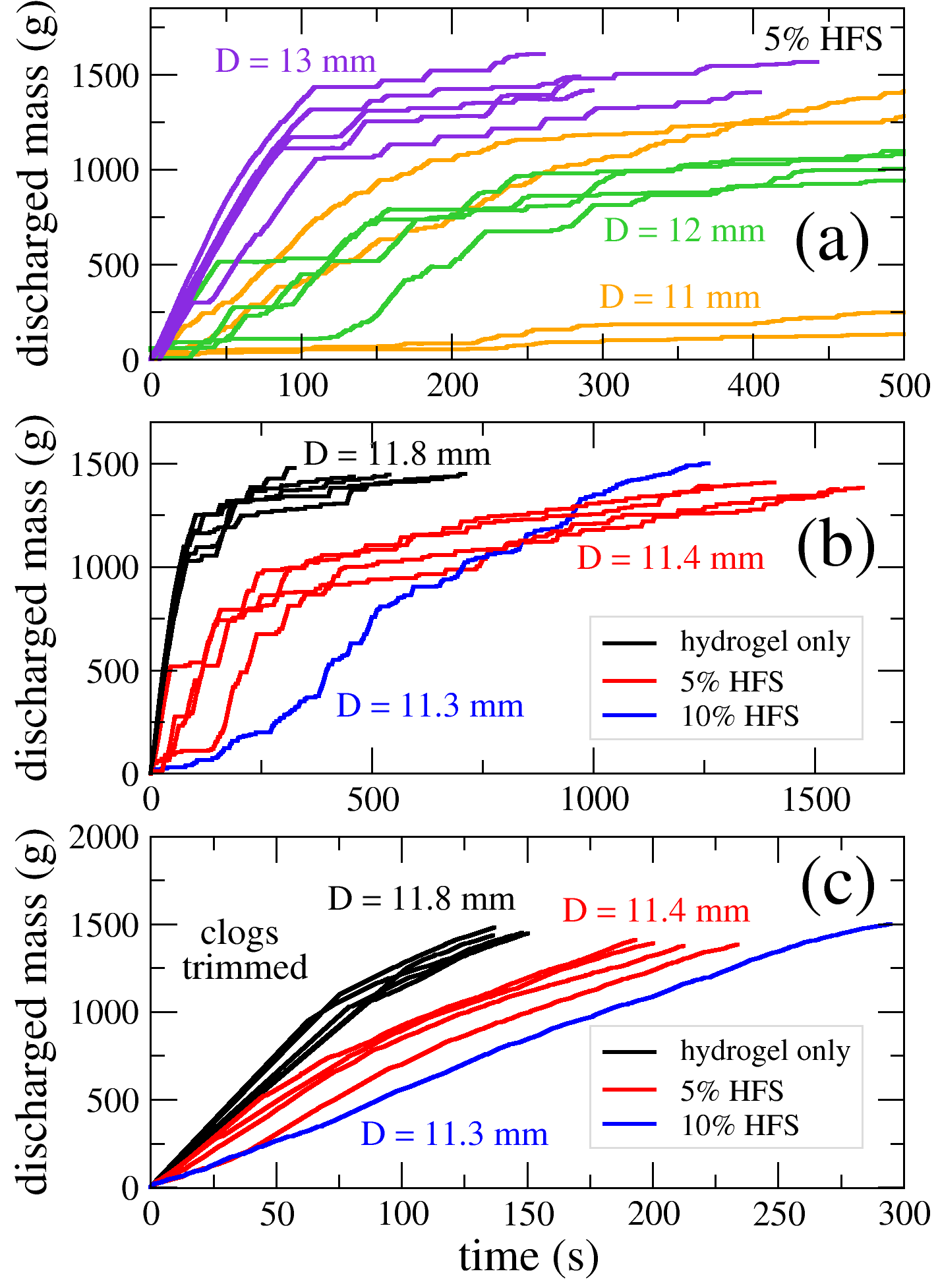}
    \caption{ (a) Time dependence of the discharged mass for a sample containing 5\% hard frictional spheres (HFS) for 3 values of the orifice width $D$.  It is evident that the probability of clogging increases strongly with lower ratios of orifice width to particle diameter. (b) Time evolution of the discharged mass  for 3 different samples: pure hydrogel, and mixtures containing 5\% or 10\% HF spheres. The orifice size $D$ is indicated on the figure. The horizontal sections correspond to the clogs. (c) Same data as panel (b) but with clogs longer than 1 s trimmed. The flow rate is measured as the local slope of these trimmed curves.}
    \label{fig:mt2}
\end{figure}

Fig. \ref{fig:mt2}a demonstrates how the character of the outflow is altered by changing the size of the orifice for a given mixture. One can see that the material flows practically uninterrupted through the largest orifice, with 13 mm width, until the fill level has lowered to about 20 ... 25 cm. In the silo with 12~mm orifice width, clogs interrupt avalanches of the order of 100~g (nearly 1000 particles), while in the silo with 11 mm opening, avalanches are on average one order of magnitude smaller.
The fill-height dependence of the occurrence of plateaus (non-permanent clogs) in the graphs is the consequence of a pressure-dependent blocking probability of individual soft grains passing the outlet, see Appendix A.



The presence of a small fraction of hard frictional particles
influences the outflow dynamics by having an impact on the
statistics of clogs (permanent or non-permanent) as well as on the
flow rate between clogs. The outflow curves are shown in Fig.
\ref{fig:mt2}(b) for 3 samples with hard frictional sphere contents $x_{\rm hf}$ of 0, 0.05 and 0.1. The  orifice width was $D\approx 11.5$ mm.  The flow rate
(between clogs) depends on the composition of the sample (fraction
of HF spheres) and other parameters, such as the filling height and orifice
size. We will further investigate the dependence of the flow rate
as well as the duration of non-permanent clogs on the number of
HF spheres in the proximity of the orifice.

\begin{figure*}[htbp]
    \centering
    \includegraphics[width=0.98\textwidth]{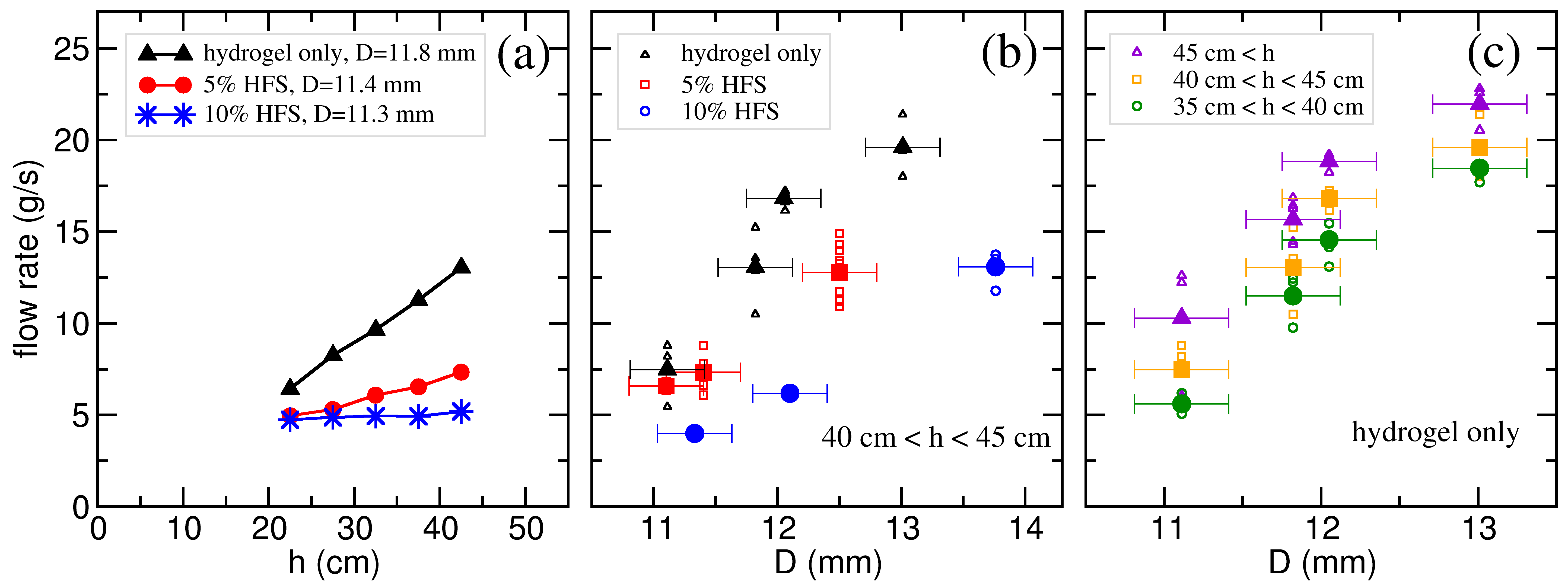}

    \caption{(a) Flow rate as a function of the bed height $h$ for pure hydrogel and for mixtures with 5\% and 10 \% HF spheres. (b) Flow rate as a function of orifice size $D$ for all 3 samples for a bed height of $40$ cm $<h<45$ cm and for (c) pure hydrogel samples at various bed heights $h$. Small open symbols represent individual experiments, large filled symbols show averages of the respective data sets (on average 4 experiments). }
    \label{fig:flowrate-D-h}
\end{figure*}

For a visualization of the evolution of the flow rate (during avalanches) during the whole discharge process, Fig.~\ref{fig:mt2}(b) was trimmed by removing all clogs longer than 1 s. The result is shown in Fig. \ref{fig:mt2}(c).
The local slope of these curves gives the instantaneous discharge rate at any moment.
This rate is shown in Fig. \ref{fig:flowrate-D-h}(a) as a function of the bed height $h$. In accordance with our earlier observations in a 3D silo \cite{Stannarius2019}, the flow rate for pure hydrogel decreases with decreasing filling height. For the sample with 5 \% HF grains, the height dependence is much weaker, while for the sample with 10 \% HF grains this trend has essentially vanished. Thus, adding a small amount of frictional hard beads to a hydrogel bead ensemble has a strong effect on the discharge kinetics of a 2-dimensional silo. Increasing the concentration of frictional hard grains, we quickly recover the typical behavior of granular materials characterized by a height independent (constant) flow rate.

It is obvious that Beverloo's original equation that relates the geometry of the particles and outlet to the discharge rate is not exact for the hydrogels and at least the 5~\% HF spheres mixture since their discharge rates depend upon pressure at the outlet.
Here, the low friction coefficient of the hydrogel may play a role \cite{Darias2020}, but the primary cause is the pressure dependence caused by the grain elasticity.
Astonishingly, addition of 10~\% of rigid grains fully removes this pressure dependence.

For an analysis how the flow rate depends on the orifice size $D$, and for comparison of all three mixtures, we present data of the height range between 40 cm and 45 cm in Fig. \ref{fig:flowrate-D-h}b. As expected, the flow rate $W$ decreases with $D$. An increased  concentration of hard grains clearly reduces the flow rate, particularly at small orifice sizes.
A dependence as
predicted by Beverloo's model in 2D, $W = C\phi \rho_0 h\sqrt{g}(D-kd)^{1.5}$,  with the grain diameter $d$, the density $\rho_0$ of the grains, packing fraction $\phi$, cell thickness $h\approx d$ and adjustable constants $k$ and $C$ may be fitted for all three
samples, but this is not surprizing because of the small $D/d$ range
and the free parameters $k$ and $C$. The product $\rho_0 \phi h$ amounts to approximately 3.7 kg/m$^2$. From Fig.~\ref{fig:flowrate-D-h}b, one finds $k\approx 1.6$, which is larger than the commonly reported value of about 1.4. The constant $C$ to
fit the graphs in Fig.~\ref{fig:flowrate-D-h}b ranges from 3.5 for the 90 \% sample to about 6 for the pure hydrogels.
This proportionality factor $C$ accounts, for instance, for details of the orifice geometry. Considering Fig.~\ref{fig:flowrate-D-h}a, one has to conclude that this factor depends upon the instant fill height, viz. the pressure at the container bottom, in the pure hydrogel and the 5 \% HF spheres samples. This is also evident from the flow rate of the pure hydrogel sample as a function of $D$ for three different height ranges, shown in Fig. \ref{fig:flowrate-D-h}(c).

\begin{figure}[htbp]
    \centering
    \includegraphics[width=0.99\columnwidth]{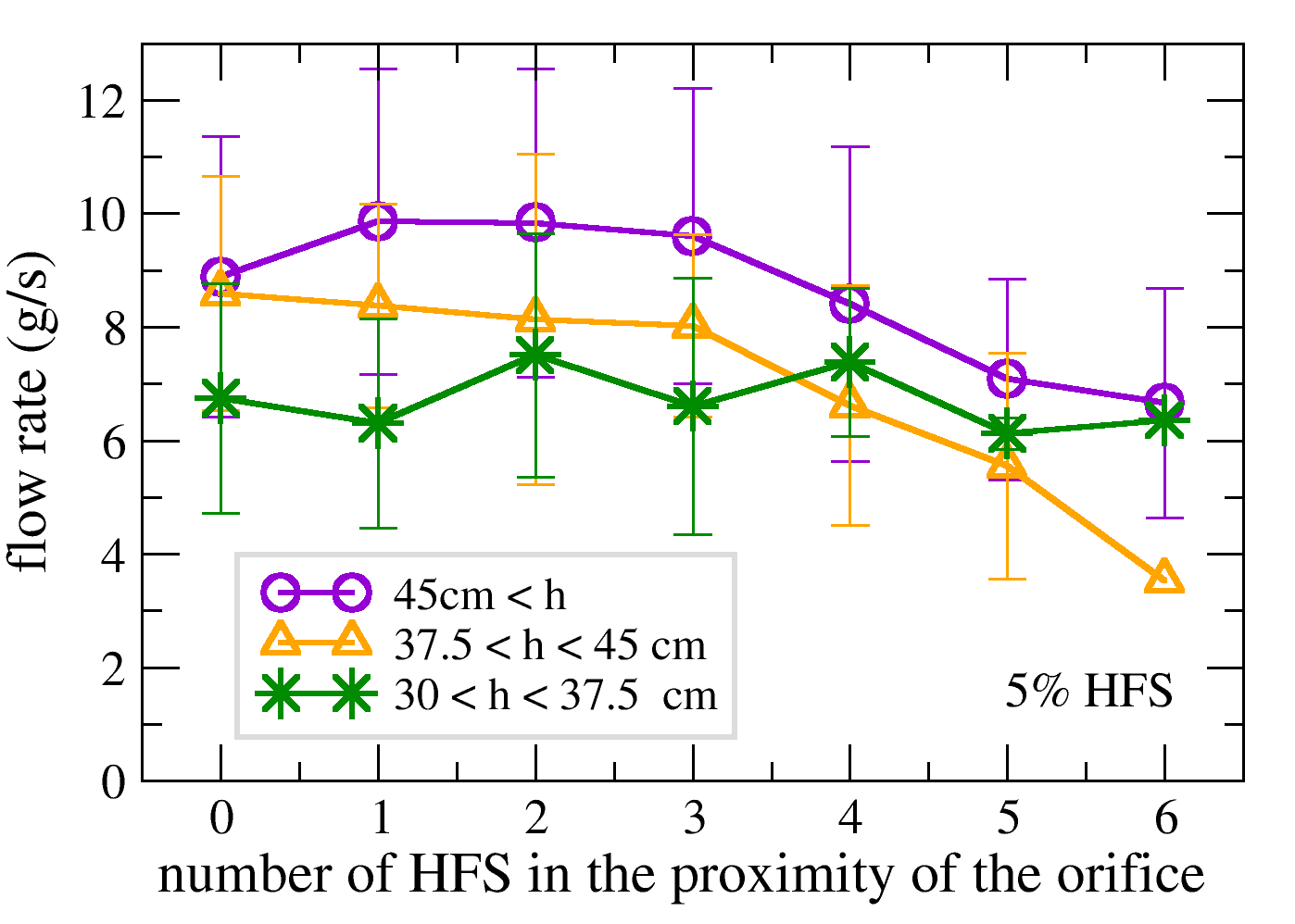}
    \caption{Flow rate as a function of the number of HF grains in the proximity \rs{(5 particle diameters)} of the orifice. Data taken for $30 ~$cm$<h<80$ cm. Data points represent 18 measurements on average, error bars stand for the standard deviation.}
    \label{fig:flow-rate-orifice}
\end{figure}

Next, we analyze the dependence of the flow rate on the number of HF grains in the vicinity of the orifice. We consider a region with the shape of a half circle with a radius of 5 $d$ above the orifice. For large bed heights ($h>37.5$ cm) the flow rate is clearly decreasing when we have \rs{more than 3} hard grains in this region, i. e. the presence of hard, frictional grains near the bottleneck has a noticeable effect on the outflow (see Fig. \ref{fig:flow-rate-orifice}). For $h<37.5$ cm, no such decrease was detected. When the pressure at the bottom is already very low, the addition of hard grains has little effect on the outflow dynamics.
The explanation is straightforward: the elasticity of the hydrogel plays a role primarily when there is high pressure at the orifice. When the silo is filled by a 40 cm high granular bed, the pressure is approximately 3 kPa, and this pressure can deform the soft particles at the orifice by roughly 10 \% of their diameter. When the fill height and consequently the pressure lowers, the deformations are much less intense and the hydrogels gradually
approach the mechanical properties of still low-frictional but hard grains.

\subsection{Non-permanent clogging}

\begin{figure}[htbp]
    \centering
    \includegraphics[width=0.99\columnwidth]{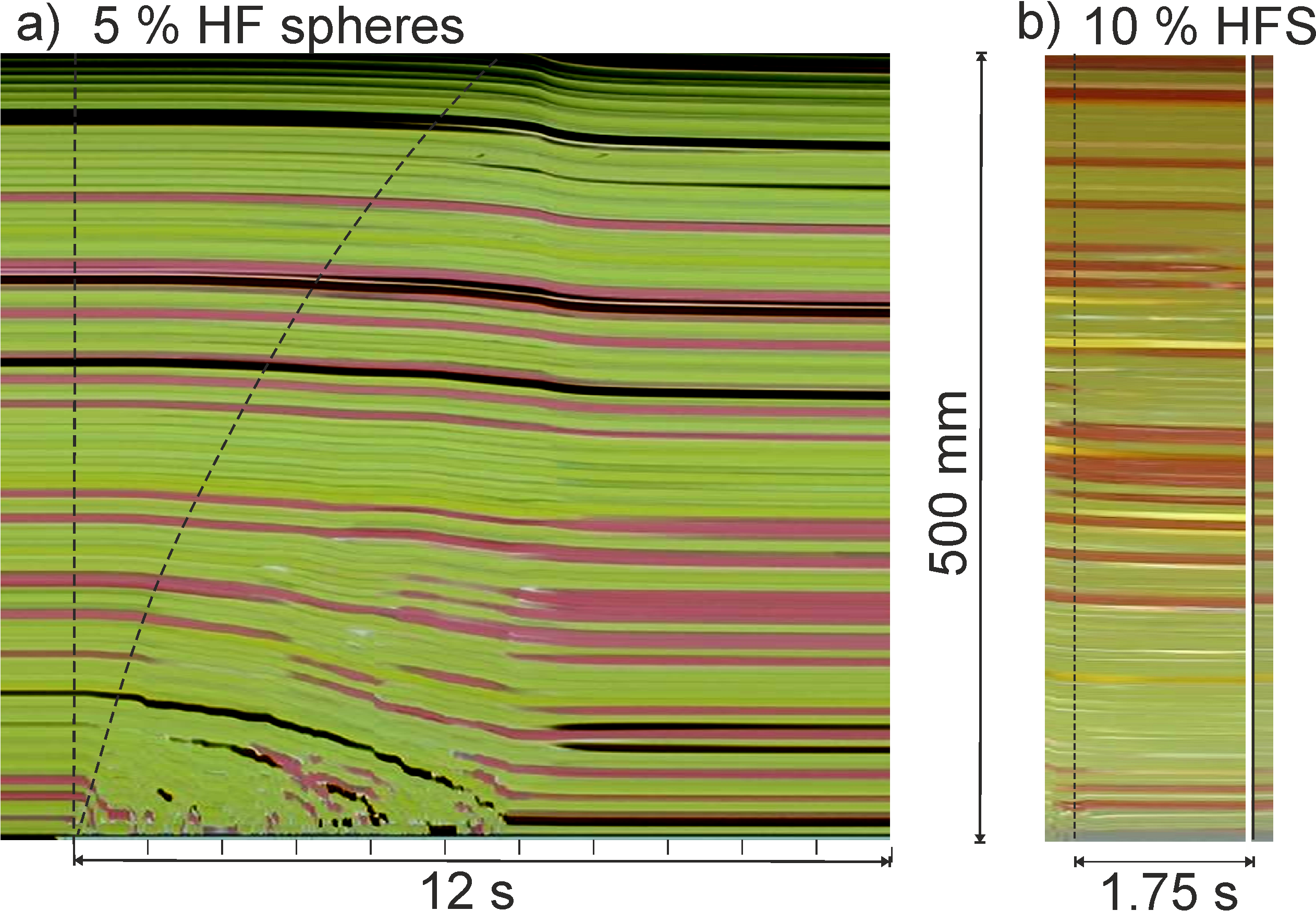}
    \caption{Space-time plots of a vertical cross section of the silo in the central axis above the orifice ($D= 12$~mm). a) Mixture with 5 \% HF spheres. The grains start to flow only locally at the orifice after it is opened (vertical dashed line). The reaction of the material in the upper part is delayed considerably (bent dashed line). b) After the orifice at the bottom clogs (dashed line), the material (with 10 \% HF spheres) in the upper part still reorganizes for several seconds. The state immediately after the outflow stopped is re-plotted behind the white gap, to visualize the changes during the clogged state.}
    \label{fig:viscoelastic}
\end{figure}

Similar to earlier observations with pure hydrogel sphere ensembles \cite{Harth2020}, the system shows non-permanent clogging. This is seen, for example, in the mass curve shown in Fig.~\ref{fig:mt}. The plateaus are signatures of stopped outflow. In the experiment shown, with 11 mm orifice size and the silo emptied to about one fourth,
roughly every second congestion of the orifice ends spontaneously, without external interference by air flushes or other. The reason for that is identified in the viscoelastic properties of the hydrogel. While the orifice is blocked, there is still motion in the upper parts of the silo that may cause an imbalance of forces in the blocking arches, with a considerable delay of up to several seconds. Figure \ref{fig:viscoelastic}b demonstrates this delay. The granular material still reorganizes in the upper regions of the container after the outflow has already stopped. During a period of 1.75 seconds, no grain leaves the orifice but the material in the upper parts rearranges slowly. The blocking structure dissolves after 1.75 seconds. We have plotted the configuration immediately after the clog started on the right hand hand side of the figure, behind the solid line. Comparison with the state at the end of the clog evidences the shift of the grains during the congestion.

The viscoelastic character of the hydrogel is the reason for the slow dynamics of these processes. It is even more evident immediately after the outlet of the freshly filled container is opened for the first time, as shown in Fig.~\ref{fig:viscoelastic}a. The flow at the orifice sets in immediately, while the motion of the grains far above the orifice is delayed by up to several seconds. Note that in silos filled with hard grains, there is practically no such delay. This phenomenon has been reported for pure hydrogel samples before \cite{Harth2020}. The delay times can vary slightly between individual runs of the experiment.
Compared to the pure hydrogel samples, the addition of few percents of hard grains
even seems to slightly enlarge the average delay times.

\begin{figure}[htbp]
    \centering
    \includegraphics[width=0.9\columnwidth]{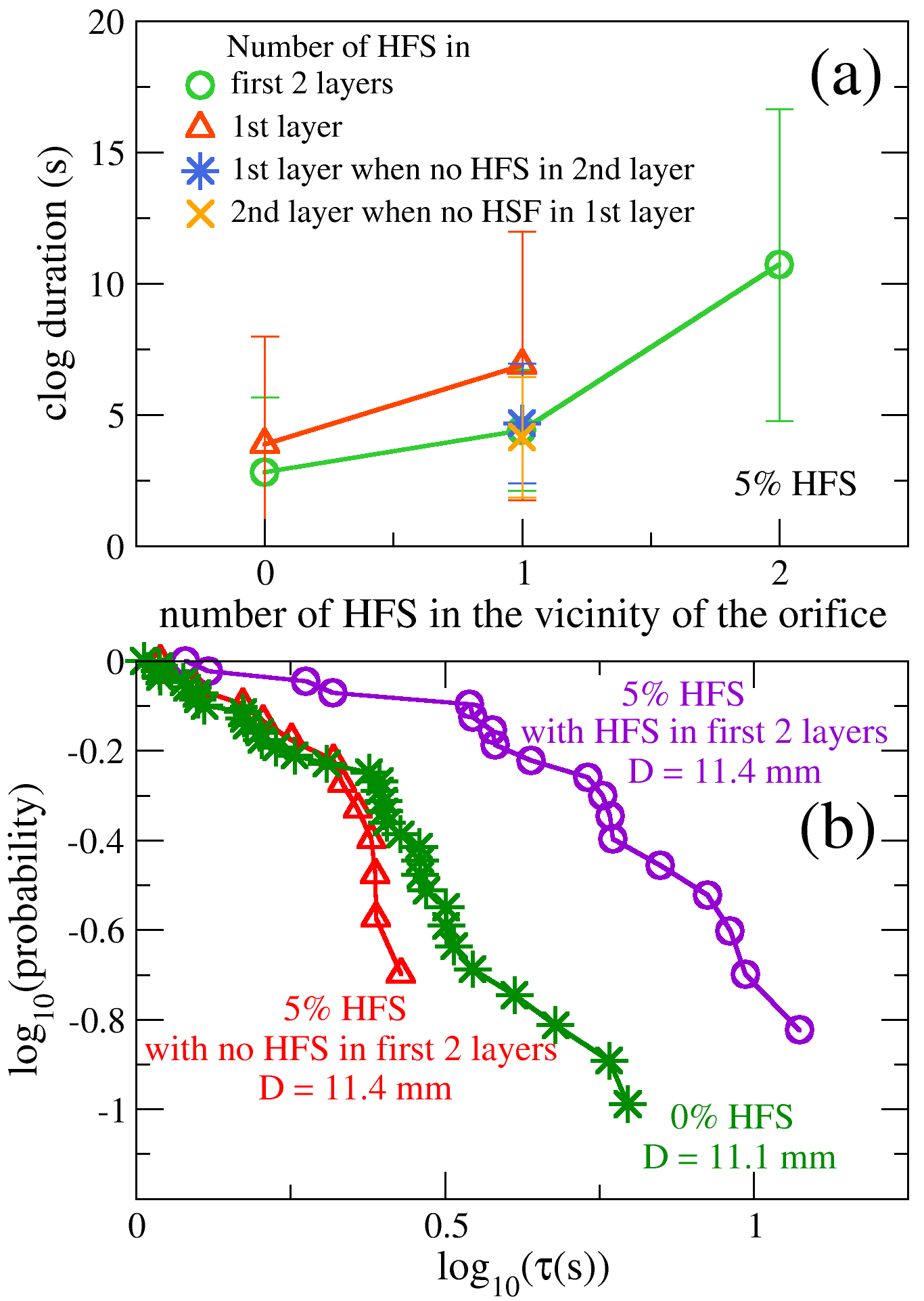}
    \caption{(a) Clog duration as a function of the number of high friction spheres (HFS) in the vicinity of the orifice. The data sets correspond  to cases when only the first layer, or the first 2 layers are considered. Data points represent 13 clogs on average, error bars stand for the standard deviation. (b) Probability of non-permanent clogs longer than $\tau$ as a function of $\tau$ on a log-log scale.}
    \label{fig:clog_duration}
\end{figure}

 We analyze now, how the duration of non-permanent clogs depends on the number of hard frictional spheres (HFS) in the vicinity of the orifice. As we see in Fig. \ref{fig:clog_duration}(a), the clog duration increases with an increasing number of HFS. Hard frictional beads in the first and second layer both have an effect on the clog duration, with a slightly larger influence of the first layer. Another way to quantify how hard frictional beads in the vicinity of the orifice influence the duration of non-permanent clogs is to plot the probability that the clog is longer than a specific time interval $\tau$. This is shown in Fig.~\ref{fig:clog_duration}(b) as a function of $\tau$ in a log-log plot. The curve characterizing a pure hydrogel sample is very close to the curve of a mixture with $5\%$ HFS, when no HFS are present in the first 2 shells above the orifice. However the curve is considerably shifted for those clogs, when HFS are present in this region. Thus, the stability of blocking arches clearly increases when hard frictional grains are present in them. This can be understood intuitively.

\subsection{Arch structure analysis}
\label{Sec:Exp:ClogStructures}

\begin{figure}[htbp]
    \centering
  \includegraphics[width=0.8\columnwidth]{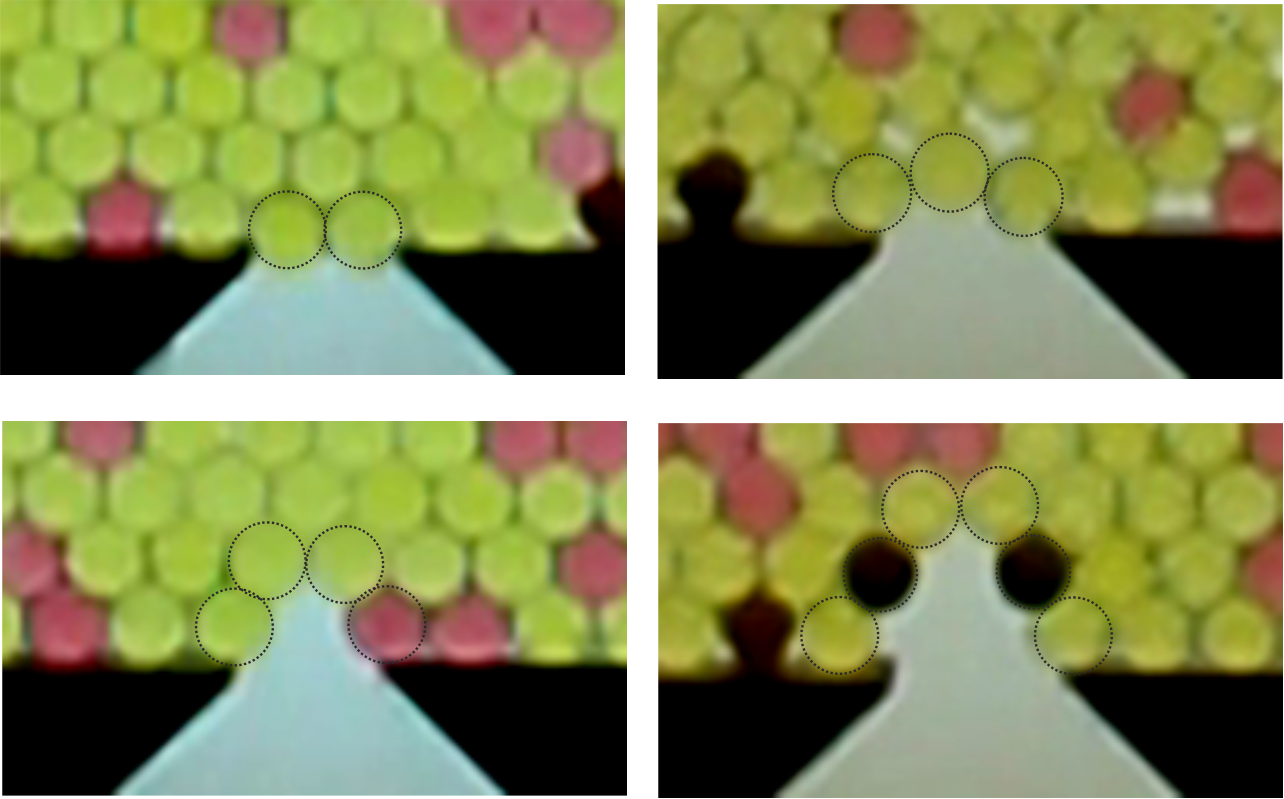}
    \caption{Typical structures of blocking arches. The 2 and 4 particle arches were taken from snapshots of clogged states of the 11~mm orifice videos, the 3 and 6 particle arches from 13~mm orifice videos.}
    \label{fig:arches}
\end{figure}

An advantage of the 2D bin with narrow orifice widths
is that only few types of clog structures are formed. First, we characterize these clogs by the number of particles involved in the first layer. The most frequently encountered structure in all three orifice sizes is the nearly symmetric four-particle arch, as seen in Fig.~\ref{fig:setup}, bottom. Also, the nearly symmetric two particle clogs are encountered more often than the other structures when the orifice width is two sphere diameters or less. Some typical structures are shown in Fig. \ref{fig:arches}.

\begin{figure*}[htbp]
    \centering
  \includegraphics[width=0.8\textwidth]{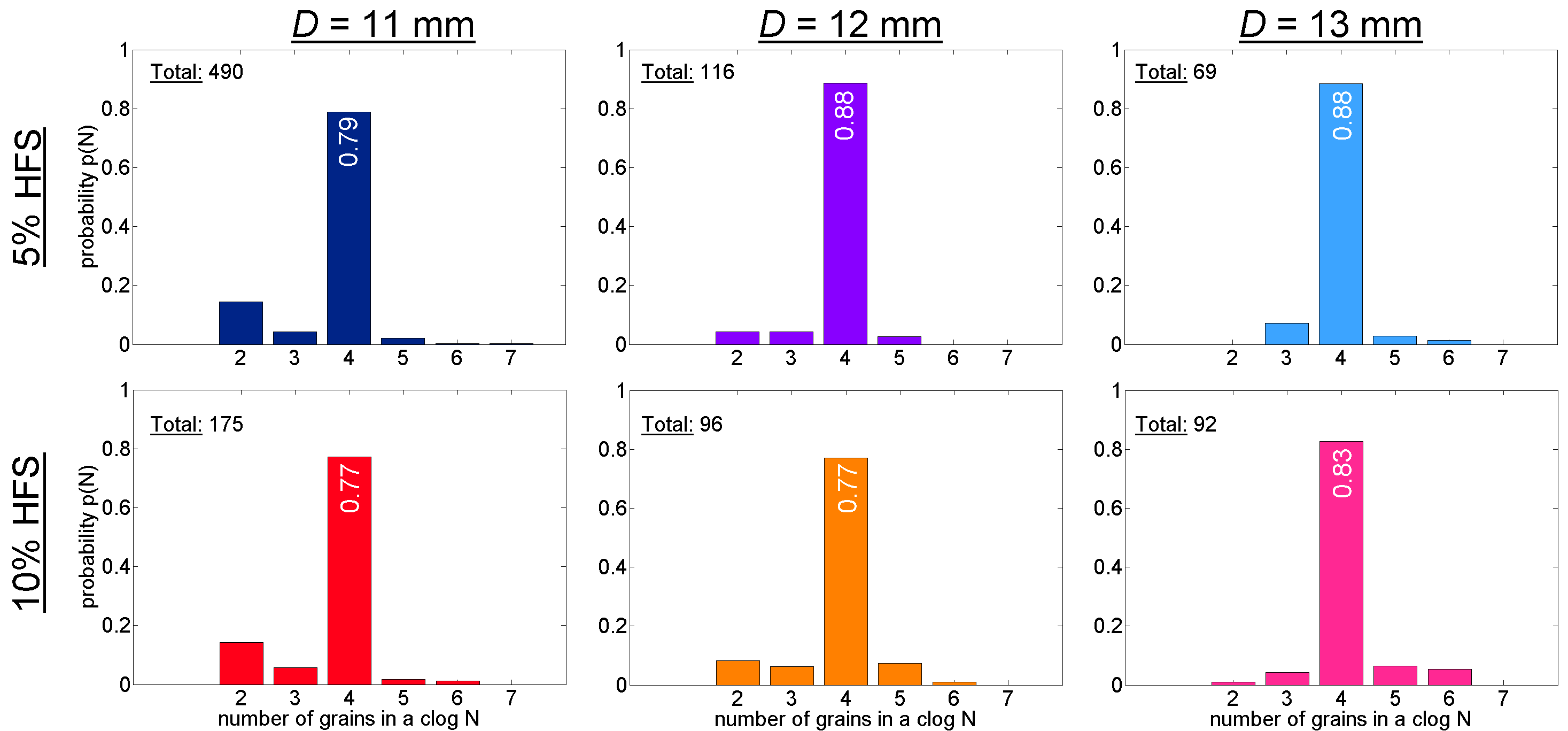}
    \caption{Number of particles forming the blocking arches at the orifice for two mixture compositions and three orifice sizes. It is seen that neither the composition of the mixtures nor the variation of the orifice width influences the dominance of 4 particle clogs. Clogs consisting of 2 particles are much less probable in the larger orifices, which is intuitively clear.}
    \label{fig:temp1}
\end{figure*}

Figure~\ref{fig:temp1} shows the statistics of clog structures grouped by the number of particles in the blocking arch for three orifice sizes and two mixture compositions. Here, we analyzed blockages that lasted 1.5 second or longer, and we did not distinguish between temporary congestions and permanent clogs that had to be destroyed by air flushes. Clogs with more than 5 particles do practically never form, three-particle clogs and five-particle clogs were found more often in the larger openings. Four-particle arches represent more than 80~\% of all blocking structures.
Therefore, we focus
primarily on the latter in our further analysis. Within the statistical error, there are no significant differences between the two mixtures.

\begin{figure}[htbp]
    \centering
    a)\includegraphics[height=0.85\columnwidth]{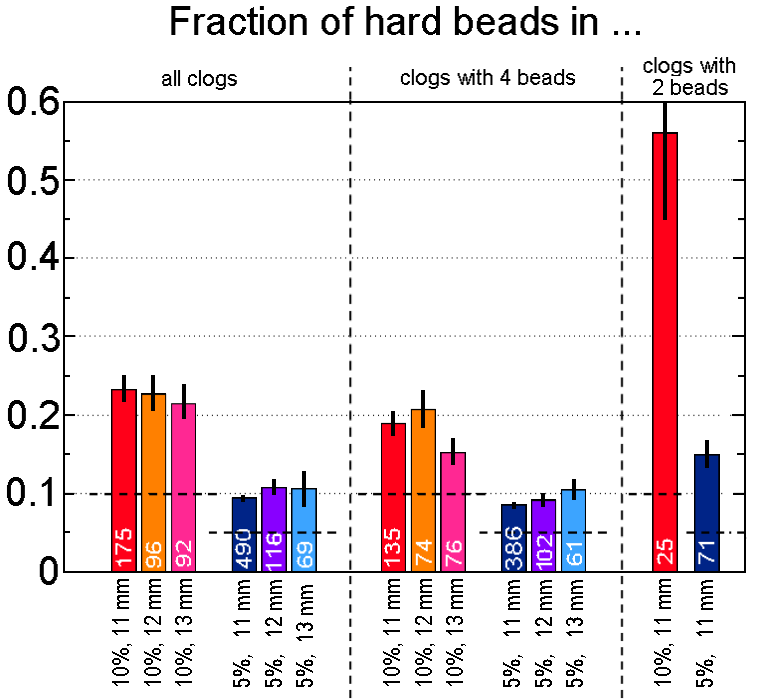}
    b)\includegraphics[height=0.7\columnwidth]{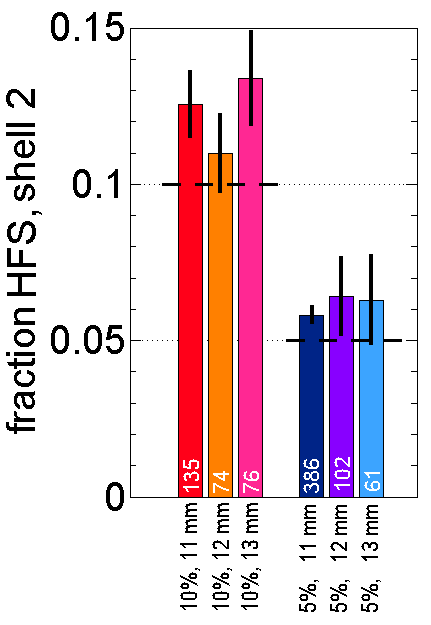}
    \caption{Statistics of the occurrence of HF spheres in the blocking arches in the experiments. The white numbers indicate the number of events compiled in the column. a) Composition of the arch, viz. the first coordination shell above the orifice; fractions $X_{\rm h}$ of hard beads compared to the bulk concentrations $x_{\rm h}$ (dashed lines). The left section includes all clogs of at least 1.5 s duration. In the middle, the statistics for 4-sphere arches is shown. On the right, data of 2-sphere arches are collected. b) Concentration of HF spheres in the second coordination shell, averaged over all clogs.}
    \label{fig:StructureStat}

\end{figure}

We will now analyze the composition of these blockages, primarily of the most frequent four sphere arches. Figure
\ref{fig:StructureStat}a shows the relative amount $X_{\rm hf}$ of hard grains in the blocking structures, separately for four sphere arches and all others. White numbers the number of arches of the respective type that occurred, summed over all evaluated experiments.
We do not distinguish here between the individual positions in the arches. The particle that finalizes the blockage may have arrived from above or from a side. Horizontal dashed lines indicate the percentage $x_{\rm hf}$ of hard grains in the mixture.
The result is that hard grains are present in the blocking arches
nearly two times more frequently than globally in the sample.
This holds, within the statistical uncertainty, for both mixtures with
$x_{\rm hf}=0.05$ and $0.1$, and for all three orifice sizes. In arches with other than four components, the HF grains are even more strongly over-represented, yet the statistical error is much larger for these numbers because of the smaller number of arches found.

One can make the simple assumption that $X_{\rm hf}$ and $x_{\rm hf}$ are
related by the approximation
\begin{equation}
\label{eq:blockage}
    X_{\rm hf} = \frac{x_{\rm {\rm hf}} p_{\rm hf}}{x_{\rm hf} p_{\rm hf} + (1-x_{\rm hf}) p_{\rm hyd} }.
\end{equation}
This equation allows us to get a rough estimate of the ratio of the blocking probabilities $p_{\rm hf}/p_{\rm hyd}$ of hard and elastic components of the mixture:
\begin{equation}
\label{eq:blockage2}
  \frac{p_{\rm hf}}{p_{\rm hyd}} =  \frac{X_{\rm hf}(1- x_{\rm hf})}{x_{\rm hf}(1- X_{\rm hf})}
\end{equation}

Taking the values of $x_{\rm hf}$ of the mixtures and a factor of $X_{\rm hf}\approx 2 x_{\rm hf}$, one obtains
${p_{\rm hf}}/{p_{\rm hyd}}\approx 2.2$. Hard grains are nearly twice as
probable to get stuck in a blocking structure at the orifice, independent of $D$ for all orifice widths studied here. It is also interesting to check how often hard spheres appear in the next layer of grains above the blocking arch. This is analyzed in Fig.~\ref{fig:StructureStat}(b). As expected, the
occurrence of hard spheres is not significantly enhanced there, the slight deviation from $x_{\rm hf}$ is within the statistical error.

With the relative frequency of $X_{\rm hf}$ of around 0.21 in four-particle arches in the 10 \% HF spheres mixtures, assuming that the occupations of site of the blocking arch are independent of each other, one would expect that
approximately 41 \% of all arches contain one hard grain, and 18 \% two hard grains.
Actually, the experimentally determined share is somewhat smaller but within the statistical error. On the other hand, in the 5 \% mixtures one has $X_{\rm hf}\approx 0.1$, which leads to expected 29 \% of
blocking arches containing a hard particle. This is in good agreement with the experiments. When one analyzes the 4-sphere arches in more detail, one finds a statistically significant larger share of HF spheres at the
two lateral positions (more than 60~\%) than at the two upper, central positions. The reason may be that the blocking by hard grains is nearly twice as effective when they are in contact with the ground plate, where they can efficiently hinder lateral motion of grains.



\section{Numerical analysis}
\label{sec:num_analysis}

\subsection{Numerical Model}
\label{sec:num}

The numerical simulations were carried out with the open source Discrete Element Method (DEM) granular simulation software LIGGGHTS \cite{kloss2012models}.
For the calculation of the inter-particle force $\vec{F}_{ij}$, a Hertz-Mindlin contact model was chosen \cite{poschel05a}, including a normal $\vec{F}^{\rm n}_{ij}$ and tangential $\vec{F}^{\rm t}_{ij}$ component, both modeled as a short-range spring dashpot interaction.
The elastic and damping particle-particle interaction coefficients are well reproduced by the model, given several input parameters, such as, the Young's modulus $Y$, the restitution coefficient $e_n$ and the friction coefficient $\mu$.
Furthermore, the Coulomb friction constraint is applied. The tangential force is cut off so that $|\vec{F}^{\rm t}_{ij}| < \mu |\vec{F}^{\rm n}_{ij}|$ is satisfied.
The used integrator resolving the particle positions and velocities, was chosen such that in a purely elastic collision, energy would be conserved. In other words it is stable.
Besides, as a simplification, we assumed the contacts between particles to be independent of each other. However, it has been shown in the past \cite{brodu2015multiple} that when soft particles are densely packed, a force model taking into account multiple contacts better captures the experimental response.
We partly compensated this effect, using the upper bound value of Young's modulus (experimental estimation), for the soft low frictional particles.

The numerical geometry mimics the experimental setup, i.~e. the width of the flat container is $40$ cm, the thickness of the cell is $6.12$ mm, only slightly larger than the particle diameter $d = 6$ mm. The two species of particles have the same size. Thus, the system is monodisperse. However, the mechanical properties of the constituents differ notably in their stiffness and friction.
For the contact between soft low-friction spheres we set a Young modulus of $Y_{\text{hyd}} = 100$~kPa and a friction coefficient of $\mu_{\text{hyd--hyd}} = 0.02$. Note that these values are consistent with earlier studies \cite{Brodu2015}. Besides, ball bouncing experiments led to a rough estimation of the hydrogels' restitution coefficient, resulting $e_{n_{\text{hyd--hyd}}} = 0.5$, approximately.

The contact between two airsoft bullets (hard, frictional) was mimicked using a Young modulus $Y_{\text{hf}} = 500 ~Y_{\text{hyd}} = 50$ MPa, $\mu_{\text{hf--hf}} = 30 \cdot \mu_{\text{hyd--hyd}} = 0.6$ and $e_n=0.8$.
In order to discriminate the effect of friction, we pursued simulations with three distinct inter-species frictions $\mu_{\text{hf--hyd}} = 0.05,\ 0.2$, and $0.3$.
The friction with the walls was set equal to the particle--particle friction i.e.~$\mu_{\text{hf--wall}} = \mu_{\text{hf--hf}}$ and $\mu_{\text{hyd--wall}} = \mu_{\text{hyd--hyd}}$, but the friction of hydrogel-like particles with the bottom plate was assumed to be similar to that with the frictional species, $\mu_{\text{hyd--bottom}} = \mu_{\text{hyd--hf}}$.
Using these parameters, we detected a maximum particle overlap of around 10\% in terms of the particle diameter. Those particles were usually found in the bottom of the system when the silo is full.

The used set of orifice widths was $D = 11, 12, 13$, and $14$ mm corresponding to aspect ratios of $\rho = 1.83, 2.0, 2.17$, and $2.33$. Similar to the experiments, we are able to observe the clogging, intermittent and continuous flow regimes using this range of orifice widths.
A simple procedure was implemented in order to resolve the occurring clogs. The number of particles exiting the container was checked every second, in case it was found to be zero,
particles located only $5$ mm away from the center of the orifice
were removed from the simulation.
Additionally, particles located in the vicinity of the hole were moved upwards within $0.1$ s,
imitating an air flush.
Simulations for each set of parameters were run four times with different random initial packings, which are taken into account in Section \ref{sec:num_analysis} in order to improve the accuracy of results.




Despite the simplicity of the interaction model, used to describe the particle-particle and particle-wall collision, the numerical results confirm that the presence of a small fraction of hard frictional particles notably impacts the outflow dynamics. The contact between different species, characterized by a friction coefficient $\mu_{\text{hyd--hf}}$, turns out to be an important parameter of the model.

\subsection{Flow rates}
\label{sec:T_flowrate}
\begin{figure}[htpb]
    \centering
    \includegraphics[width=0.48\textwidth]{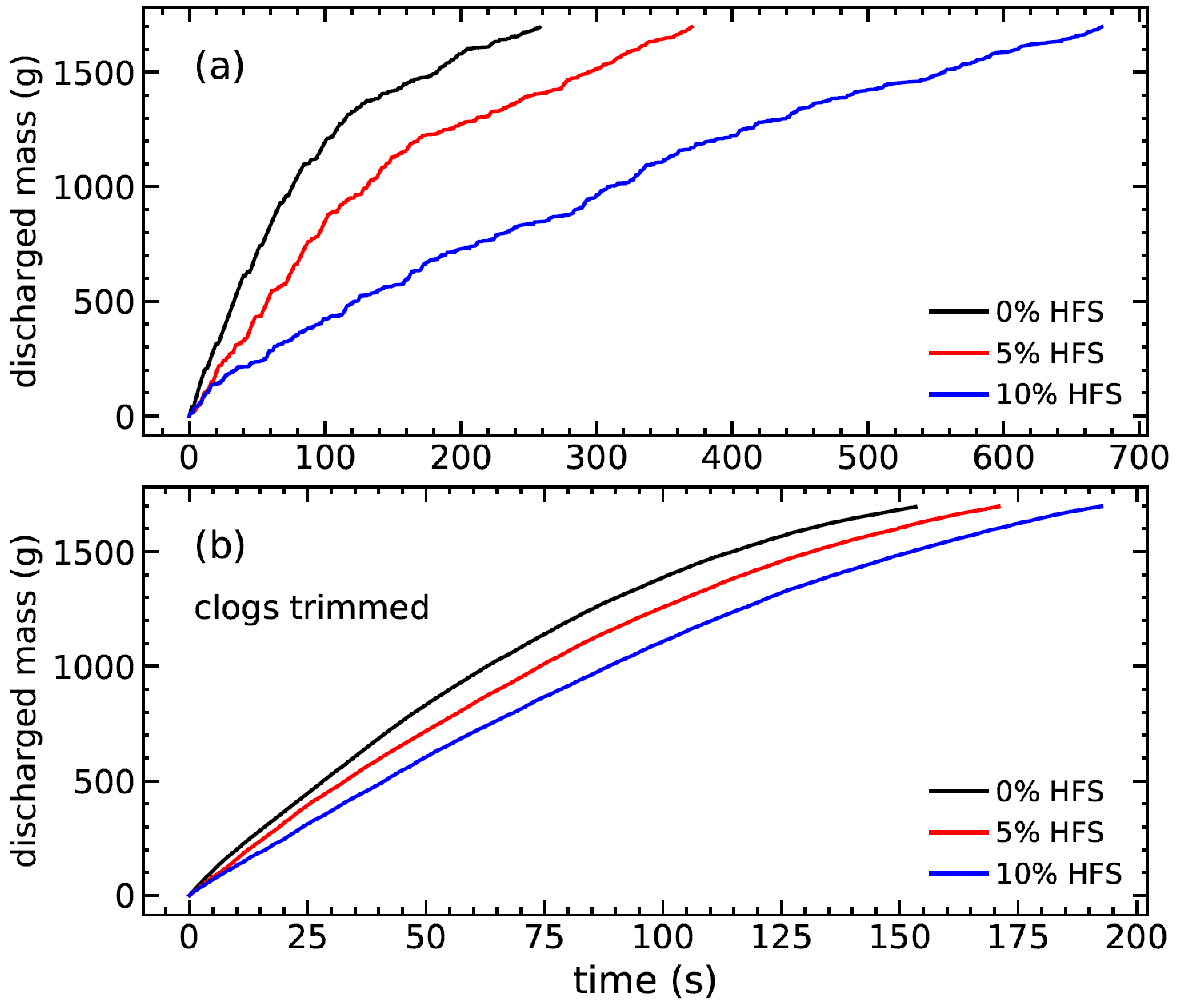}
    \caption{Discharge characteristics calculated numerically for $D = 11$ mm, $\mu_{\text{hyd--hf}} = 0.2$. The top graph shows the original data, the bottom graph presents the same data adjusted by trimming the phases of stopped outflow.}
    \label{fig:sim_mass_vs_time}
\end{figure}

In a first step, we examine the outflow as a function of the mixture composition, fixing the size of the orifice and using $\mu_{\text{hyd--hf}} = 10\, \mu_{\text{hyd--hyd}}$. Figure \ref{fig:sim_mass_vs_time}(a) displays results obtained for a system containing only soft, low-friction particles, and systems with 5\%, and 10\% of hard frictional particles, the remaining spheres being soft, with low-friction. As noticed, the addition of even a small number of hard frictional grains leads to a significant reduction of the mass discharged for the same period of time. The homogeneous system of soft, low-friction particles discharges much faster while adding hard frictional particles induces the formations of clogs, where the particle flow is interrupted randomly.  Moreover, similar to the experimental scenario, both the frequency and the duration of the clogs increase with the increasing fraction of hard, frictional particles.
For clarity, Fig.~\ref{fig:sim_mass_vs_time}(b) also shows the analysis of the mass vs. time curves, after removing the periods of time in which the flow rate is zero (between avalanches). These data allow us to numerically obtain the evolution of the mass flow rate $ {dm}/{dt}$ in time, as well as to analyze its dependence on the filling height and orifice size.

\begin{figure}[htpb]
    \centering
    \includegraphics[width=0.8\columnwidth]{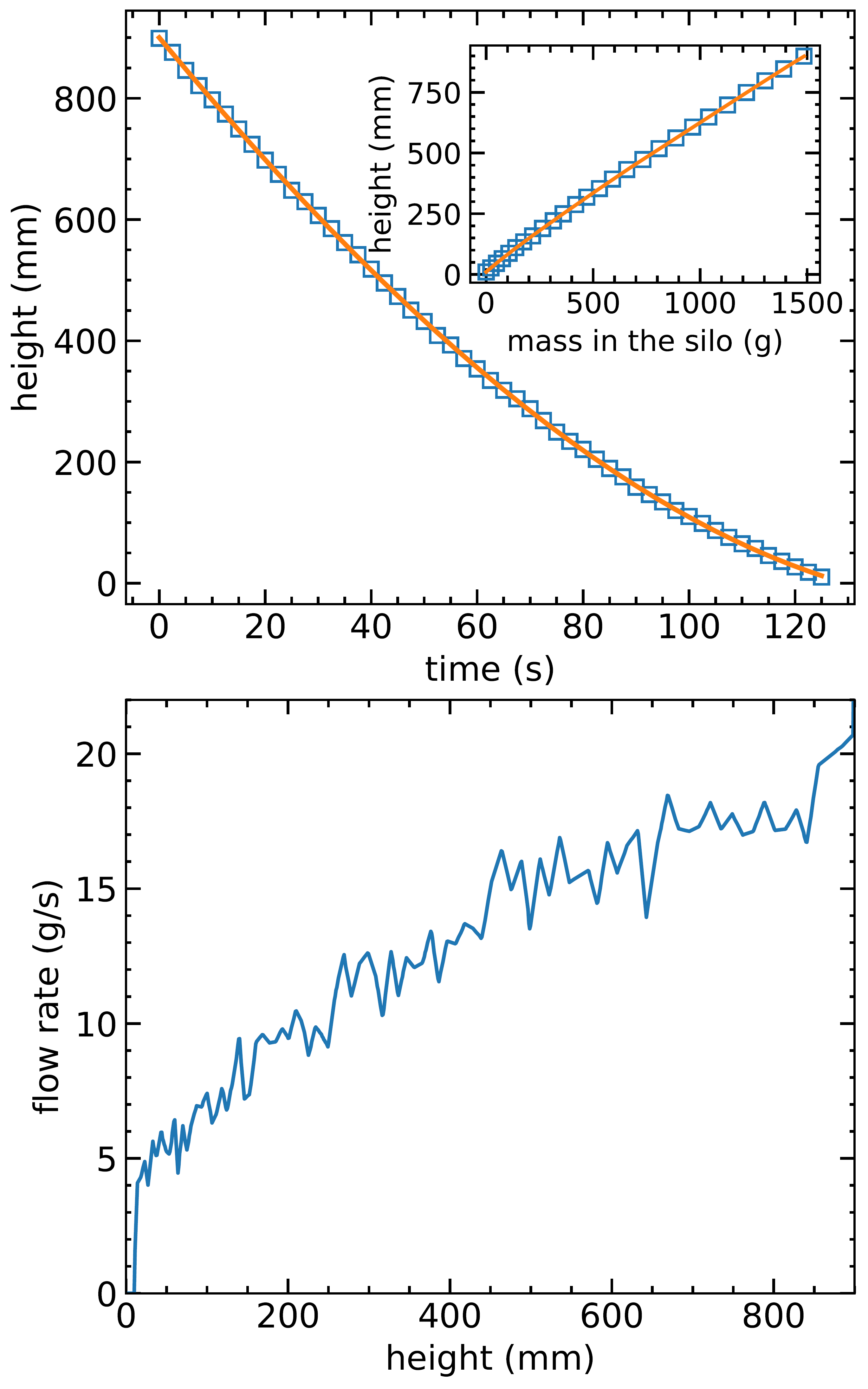}
    \caption{Discharge characteristics calculated numerically for $\mu_{\text{hyd--hf}} = 0.05$, $D = 12$ mm, 5\% hard spheres. The top image shows the granular bed height $h(t)$ using the trimmed data, the bottom image gives the discharge rate as a function of the height. Small statistical fluctuations result from the computation procedure, while the trend of decreasing flow rate with lower bed height is systematic. }
    \label{fig:sim_flowrate_introduction}
\end{figure}

For further analysis, it is necessary to accurately resolve the time evolution of the granular bed height. The mean bed height $h(t)$ is obtained by sampling ten equally sized vertical slices, and locating the highest particle $h_k$ in each of them. Then, the value of $h(t)$ is found as the average of that set. For this purpose, we use the trimmed data as shown in Fig.~\ref{fig:sim_mass_vs_time}b, thus $h(t)$ is a function of the trimmed time.
Figure \ref{fig:sim_flowrate_introduction}a exemplifies results of this procedure for a system with 5\% of hard frictional particles $\mu_{\text{hyd--hf}} = 2.5~\mu_{\text{hyd--hyd}}$, and orifice diameter $D = 12$~mm.
The main figure shows the height versus time behavior, while the inset illustrates the evolution of the height versus mass remaining in the container during discharge. The latter slightly deviates from a straight line due to the compression of particles, contrary to usual hard granular media where the height is approximately linearly proportional to the mass in the silo. In addition, the dependence of the mass flow rate $\frac{dm}{dt}$ on the column height can also be deduced numerically, see for example Fig.~\ref{fig:sim_flowrate_introduction}(b).

\begin{figure*}[htpb]
    \centering
    \includegraphics[width=1.0\textwidth]{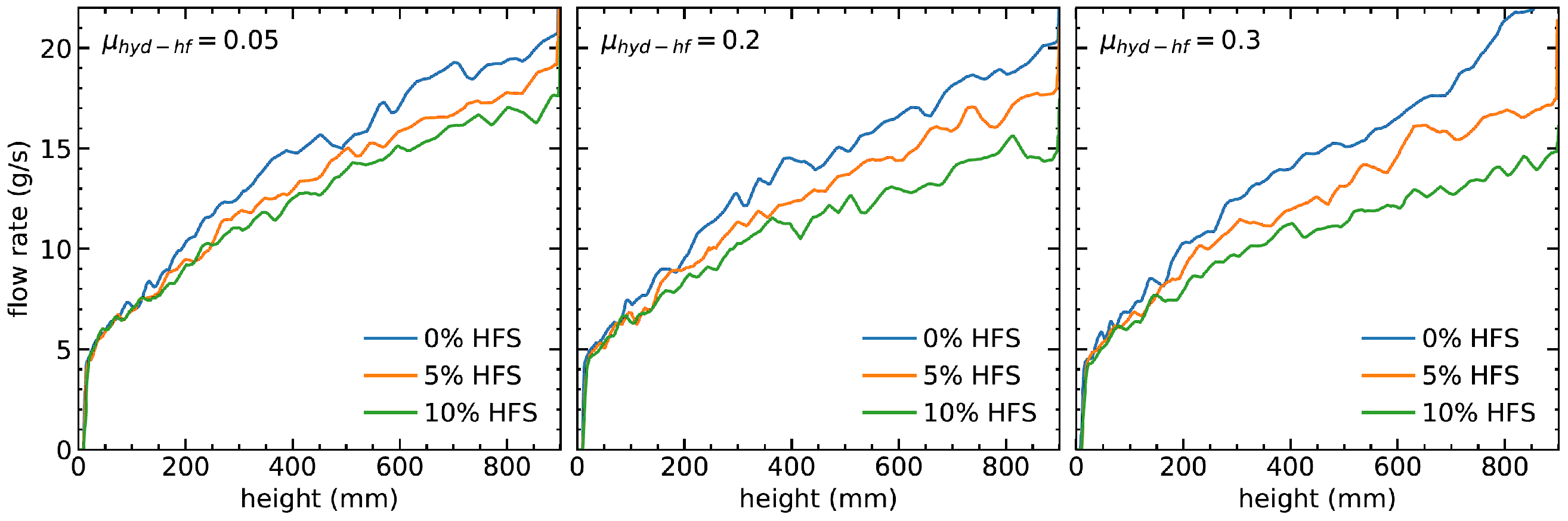}
    \caption{Numerically calculated flow rate as a function of bed height for different mixture ratios and an orifice size of $D = 12$ mm. The interspecies friction i.e. the friction between the HF and the soft, low-friction particles, reflected in $\mu_{\text{hyd--hf}}$, is different in the subfigures.}
    \label{fig:sim_flowrate_vs_height}
\end{figure*}

Figure \ref{fig:sim_flowrate_vs_height} summarizes our systematic study of the mass flow rate $\frac{dm}{dt}$ versus height $h$, performed numerically.
Specifically, we executed simulations varying the system composition and the friction coefficient between both particle types, $\mu_{\text{hyd--hf}}$, keeping the orifice diameter of $D=12$~mm constant.
In general, the numerical outcomes reflect similar trends as the experimental ones,
(see Sec.~\ref{sec:flowrate}).
First, the numerical model reproduces that the mass flow rates for mixtures and for the homogeneous system of soft, low-friction particles strongly decrease with decreasing filling height in the experiment.
In addition, we also detect numerically that this trend is affected significantly when the mixture composition changes:  already a small amount of hard frictional particles notably influences the discharge process.
Furthermore, as the concentration of hard frictional particles is increased,
the changes in flow rate become weaker, and the system tends to recover the ''classical granular response'' of a height-independent mass flow rate.

Additionally, we numerically explore the impact of the inter-species friction coefficient
$\mu_{\text{hyd--hf}}$ on the results, keeping $\mu_{\text{hyd--hyd}}$ and $\mu_{\text{hf--hf}}$
constant. Figures \ref{fig:sim_flowrate_vs_height}(a), \ref{fig:sim_flowrate_vs_height}(b) and
\ref{fig:sim_flowrate_vs_height}(c) illustrate outcomes corresponding to
$\mu_{\text{hyd--hf}} = 2.5  \mu_{\text{hyd--hyd}}$,
$\mu_{\text{hyd--hf}} = 10 \mu_{\text{hyd--hyd}}$,
and $\mu_{\text{hyd--hf}} = 15 \mu_{\text{hyd--hyd}}$,
respectively. First, it is obvious that the value of $\mu_{\text{hyd--hf}}$
significantly impacts the magnitude of the mass flow rate $\frac{dm}{dt}$. Besides, the impact of the hard frictional particles is enhanced, as $\mu_{\text{hyd-hf}}$ increases. Thus, the system's response approaches a height independent mass flow rate.

\begin{figure*}
    \centering
    \includegraphics[width=1.0\textwidth]{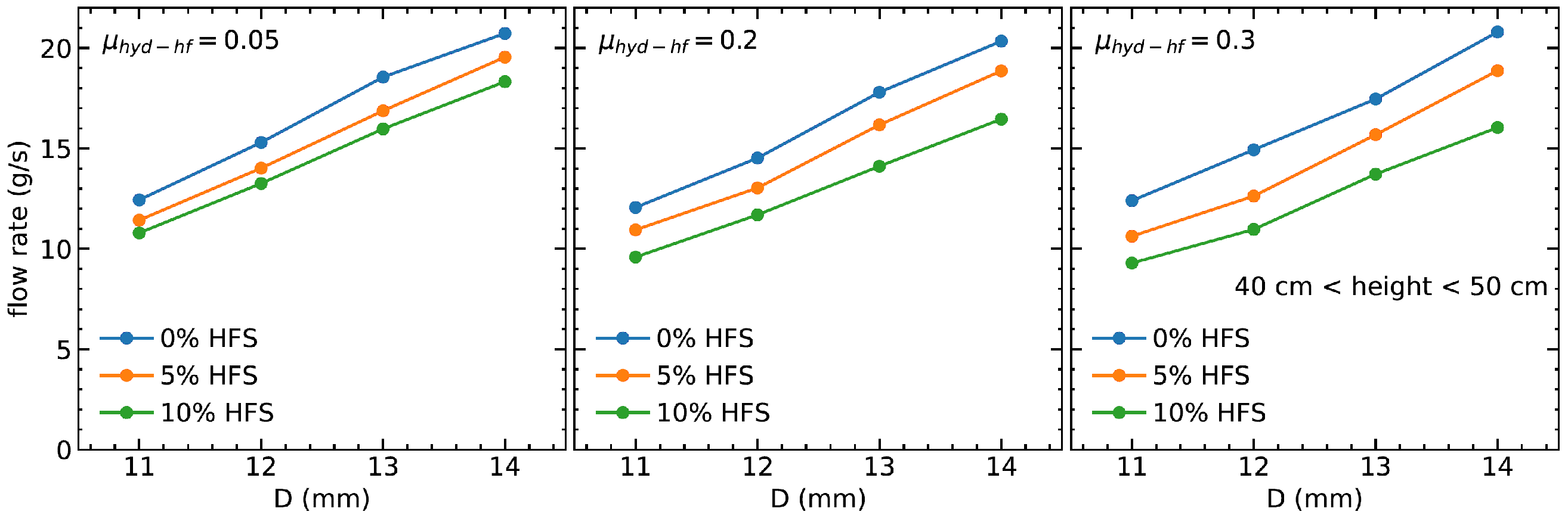}
    \caption{Average flow rate in a small range of bed height ($40$ cm $< h <$ $50$ cm) as a function of orifice size. In these numerical simulations, the inter-species friction is increased from the left figure to the right $\mu_{\text{hyd--hf}} = 0.05;~0.2;~ 0.3$.}
    \label{fig:sim_flowrate_vs_D}
\end{figure*}

Figure~\ref{fig:sim_flowrate_vs_D} shows the mass flow
rate obtained for different orifice diameters. For clarity, the calculations have been
performed with constant $\mu_{\text{hyd--hyd}}=0.02$ and $\mu_{\text{hf--hf}}= 30~ \mu_{\text{hyd--hyd}}$ and
exploring the impact of $\mu_{\text{hyd--hf}}$ on the results.
The presented data correspond to averaged values over different time intervals, in terms of the
displacement of the granular bed's surface. As expected, we detect that the mass flow rate increases
with increasing orifice diameter. Similar to the experimental results, the data
can be approximated in this narrow range by a linear increase. The specific values of the flow rate decrease significantly with the introduction of hard frictional particles. In addition, we also observe that the
friction coefficient between particles of different types significantly affects the outflow
dynamics. Specifically, the flow rate is notably reduced when increasing
$\mu_{\text{hyd--hf}}$. This suggests that stable arches composed of different particle types
play a significant role in the outflow dynamics. It is plausible that the increase of
$\mu_{\text{hyd--hf}}$ enhances the stability of mixed arches, which is even noticed when
$\mu_{\text{hyd--hf}}= 2.5 \mu_{\text{hyd--hyd}}$.

\subsection{Pressure characteristics}
\label{sec:T_pressure }

\begin{figure}[htpb]
    \centering
    \includegraphics[width=0.5\textwidth]{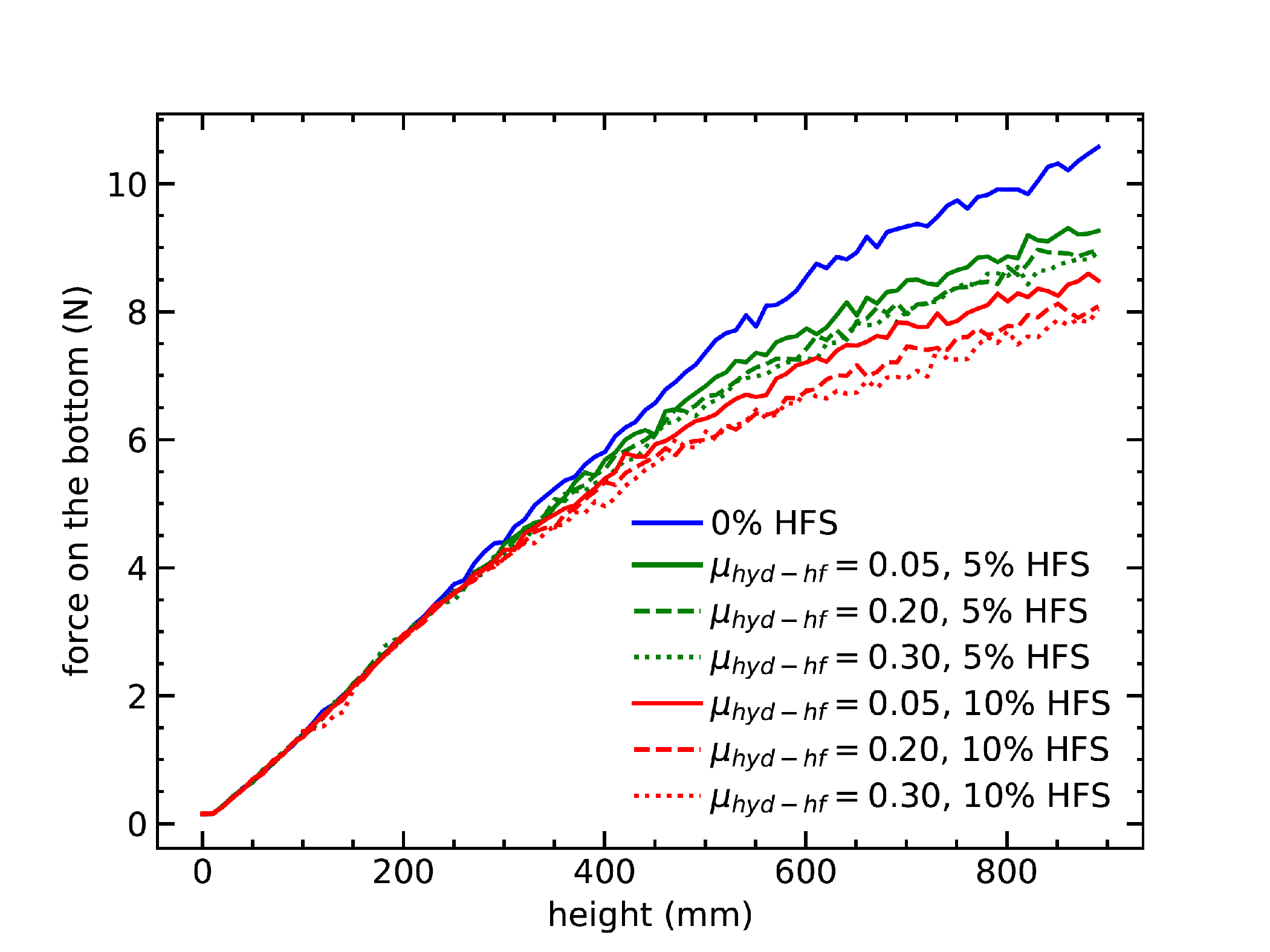}
    \caption{Force acting on the bottom plate as a function of the bed height (numerical data, orifice size $D = 12$ mm; cf. experimental data in Fig. \ref{fig:pressure}).}
    \label{fig:sim_force_vs_height}
\end{figure}

Section \ref{sec:flowrate} and Ref.~\cite{Harth2020} include experimental results which indicate that changes of the
mass flow rate with the height correlate with changes of the pressure at the bottom of the
system.  Figure~\ref{fig:sim_force_vs_height} illustrates the numerical outcome of the pressure at the bottom of the silo, obtained for several mixture compositions and particle frictions $\mu_\text{hyd--hf}$. Note that for a homogeneous system composed of low-friction particles, the curve is very close to hydrostatic behavior (linear increase). The observed deviation is consistent with the low but still nonzero friction coefficient ($\mu_\text{hyd--hyd}=0.02$) of the particles.
Similar to the experimental findings, the pressure at a given time (or granular bed height) changes significantly in the simulations if hard frictional particles are added, and the system response deviates from hydrostatic behavior. This shows that the presence of hard and frictional particles induces scattering of the stress transmission with respect to the direction of gravity. As a result, the stress transmission to the container walls is enhanced. Once again, we find that the value of $\mu_{\rm hyd-hf}$ is highly relevant, as the previously described trend enhances with increasing $\mu_{\rm hyd-hf}$.

\begin{figure}[htpb]
    \centering
    \includegraphics[width=0.5\textwidth]{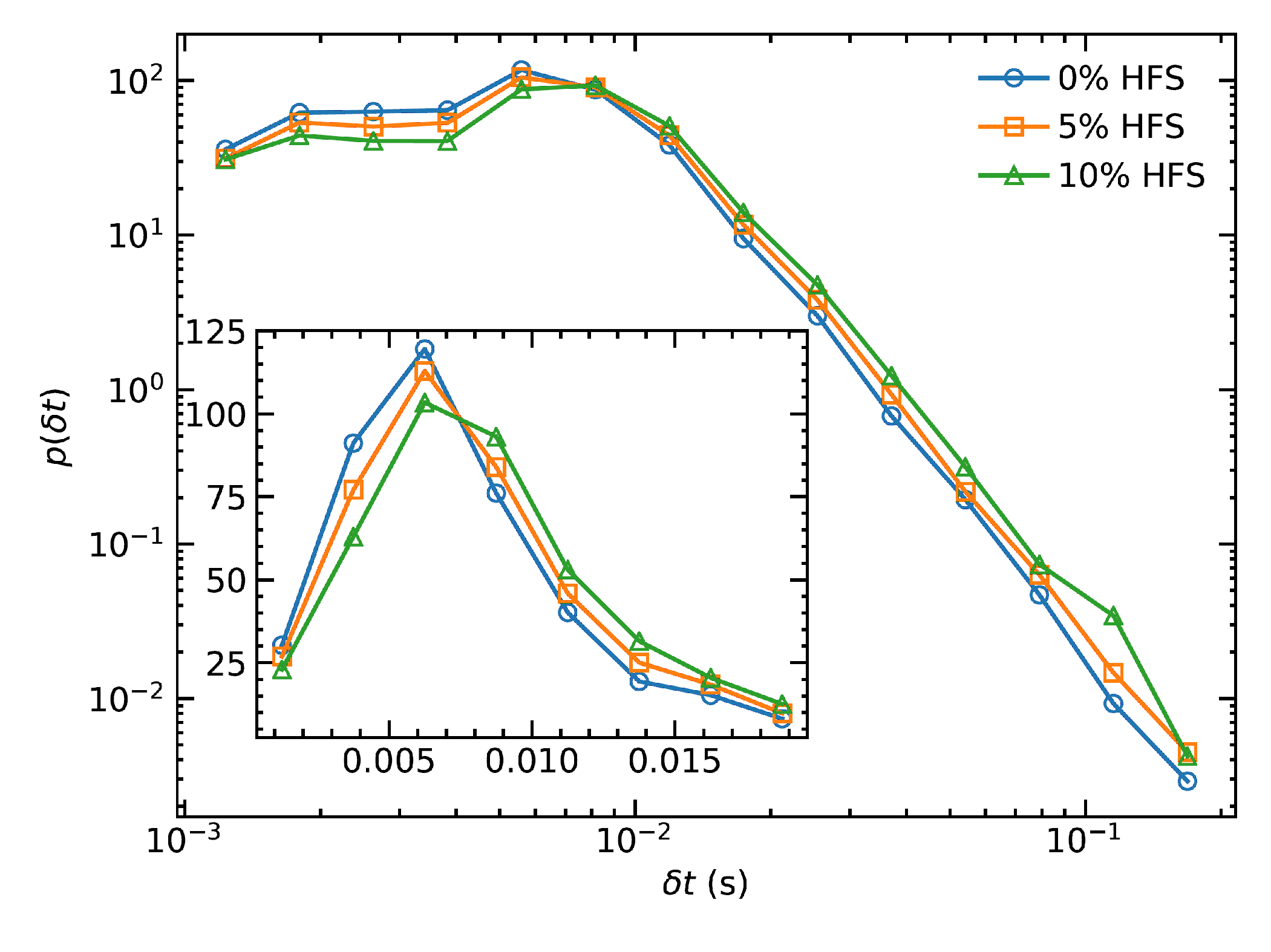}
    \caption{Probability density of the elapsed time between two particles passing in log-log representation. In these numerical simulations, the inter-species friction was $\mu_{\text{hyd--hf}} = 0.2$, while the orifice size was $D = 12$ mm. The inset shows the same data but in a smaller region with linear axes.}
    \label{fig:sim_time_between_pts}
\end{figure}

\subsection{Flow intermittency and structure of the arches}
\label{sec:T_passingtimes}

The statistics of the passing times $\delta t$, defined as the time lapse between the passing of two consecutive particles through the orifice also reveals interesting features of the clogging process. In Fig.~\ref{fig:sim_time_between_pts},
we illustrate the probability distributions $p (\delta t)$ obtained in systems with different
compositions for comparison. The first issue is the notable presence of very fast events, which are less likely when a small amount of hard frictional particles is added. Similar to the experimental trend of the clogging statistics, the tail of the distribution also gets slightly fatter if one increases the percentage of hard grains in the mixture. The distribution also shows a well defined peak at $t_d \approx 0.006$\,s, regardless of the inter-species friction $\mu_{\text{hyd--hf}}$ (see Fig.~\ref{fig:sim_time_between_pts} inset). In those specific cases, we find that the most probable vertical velocity of particles crossing the orifice is $v_d\approx 0.55$\,m/s, significantly larger than the corresponding {\it free fall} value $\sqrt{2gR}$. This numerical result indicates that the up-down pressure gradient at the orifice region notably affects the particle outflow. Moreover, it also suggests that on average two consecutive particles leave the silo within a vertical distance of approximately $D/2$.

\begin{figure*}[!htbp]
    \centering
    \includegraphics[width=0.9\textwidth]{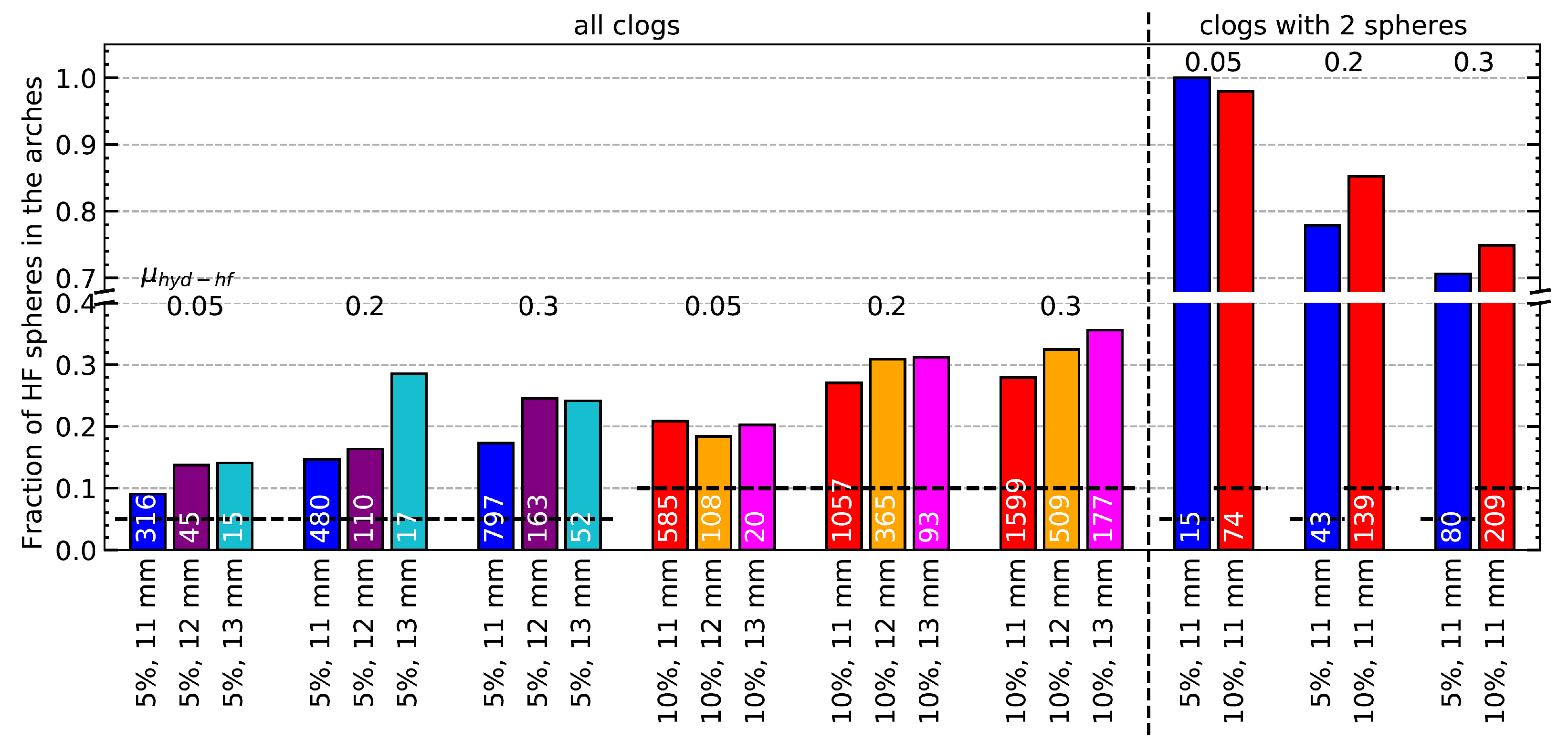}
    \caption{Statistics of the fraction of HF spheres in the blocking arches in clogged states in numerical simulations. The bars represent the mean fraction of HF spheres in the arches compared to the bulk concentration (dashed horizontal lines). At the right of the vertical dashed line, the clogs composed of only 2 spheres are considered, while on the left all occurrences are included. Under each bar, the percentage of HF spheres in the silo and the orifice size are indicated. The inter-species friction $\mu_{\text{hyd--hf}}$ is shown. The white numbers indicate the number of events compiled in the column (cf. Fig. \ref{fig:StructureStat} for experimental data)}.
    \label{fig:sim_clog_statistics from simulations}
\end{figure*}

 After identifying the characteristic time of the passing times' statistics, we here define an infinite passing time when $\delta t = 1$~s, for practical reasons. After that, the system is considered permanently clogged, and a stable arch blocks the orifice.
Fig.~\ref{fig:sim_clog_statistics from simulations} illustrates the statistics of the occurrence of hard frictional spheres in the blocking arches. One of our main findings is that in all cases, the fraction of hard frictional particles forming arches was higher than their bulk fraction, indicated by the horizontal dashed line.  We achieved the closest match to the experimental results for all clogs when an inter-species friction  $\mu_{\text{hyd--hf}} = 0.05$ was chosen.
Seemingly, the larger the orifice size is, the more frequently hard frictional particles occur in arches, while the number of blocking events decreases.
Consistently, the increasing inter-species friction also enhances the occurrence of hard frictional spheres in arches considering all clogged states, which again shows its relevance. To gain better insight, we separately examined clogged states with arches composed of only two particles: In these cases, the arches are overwhelmingly often made up of two hard frictional spheres. Moreover, increasing the inter-species friction decreases the occurrence of hard particles in the arches. This effect can be easily explained by the transition from arches composed of two hard spheres to the configuration with one hard and one soft sphere.

\section{Discussion and Summary}

Our results clearly show that the addition of a small percentage of hard grains to an ensemble of soft, low-friction spheres has dramatic consequences for the flow through a narrow orifice and the clogging statistics in a quasi-2D silo. First, one finds that the outflow rate of pure soft grain ensembles through narrow orifices depends upon the pressure at the silo bottom, which is in contrast to Beverloo's model for hard grains. Second, one observes that the addition of a few percent of hard spheres restores the pressure-independent outflow characteristics predicted by Beverloo (Fig.~\ref{fig:flowrate-D-h}(a)). Moreover, the discharge rates of all mixtures approach each other at low container fill heights (Fig.~\ref{fig:flowrate-D-h}a). The reason is obviously that the elasticity of the soft grains is less important when the pressure near the orifice is low, then their deformability can be neglected. On the other hand, at high fill levels of the silo, the increased pressure at the bottom can deform the soft grains and squeeze them through the orifice efficiently. Therefore, the outflow rate at a fill height of 40 cm (pressure $\approx 3$ kPa) is approximately 2.5 times higher for the pure hydrogel sample than for the 10\% mixture. This cannot be explained by the smaller
friction coefficient \cite{Darias2020} of the hydrogel compared to the hard frictional spheres, since the concentration of the latter is small.
The effect of the number of hard frictional grains in the vicinity of the orifice on the flow rate is in accordance with the above described observations. At higher fill levels ($h>37.5$ cm), the flow rate is found to decrease when the number of HFS is larger than 3 in the proximity (5 $d$) of the orifice (Fig.~\ref{fig:flow-rate-orifice}). At low fill levels, when the pressure near the orifice is low, increasing the number of HFS around the orifice does not affect the flow rate significantly.


We note that a comparison of outflow rates with those of a pure hard sphere ensemble is not possible because the latter will permanently clog at orifice widths smaller than $2 d$, with mean avalanche sizes   well below 10 particles.

Even though the probability of two hard grains reaching the orifice simultaneously and blocking the outlet is very low,
we find a substantial influence of hard frictional particle doping on the clogging statistics, both in experiment and simulations. The probability that a hard frictional particle is involved in the formation of the blocking arch is, on average, twice as large as for soft hydrogel spheres in the mixtures. However,
the time evolution shown in Fig.~\ref{fig:mt2} shows that
the effective discharge is considerably delayed by frequent
intermittent stagnations of the flow in the doped mixtures.

The material forms non-permanent clogs which are resolved after some delay, because of slow reorganizations of the packing structure in the granular bed within the complete container above the orifice.
This reorganization occurs on a timescale of a few seconds, as illustrated by Fig. ~\ref{fig:viscoelastic}(a). Interestingly, the probability distribution of clog durations is very similar for a pure hydrogel sample and a 5$\%$ mixture if no HFS are present in the first two layers above the orifice (Fig. \ref{fig:clog_duration}(b)). The clog duration clearly increases when HFS are present near the orifice. This observation underlines the important role of hard frictional beads in the vicinity of the orifice.

Numerical simulations also shed light on the impact of the inter-species friction on the dynamic behavior of the granulate. For instance, the deviations in the flow rate between the different ensembles were enhanced when increasing $\mu_{\text{hyd--hf}}$  (see Fig.~\ref{fig:sim_flowrate_vs_height} and Fig.~\ref{fig:sim_flowrate_vs_D}). Moreover, the total vertical force acting on the bottom wall is also slightly affected by this parameter (see Fig.~\ref{fig:sim_force_vs_height}). In general, increasing the friction between the two types of particles leads to more frequent clog events, and in those, the occurrence of HFS in the blocking arches is favored, significantly (see Fig.~\ref{fig:sim_clog_statistics from simulations}).
One explanation of the much higher importance of $\mu_{\rm hf-hyd}$ as compared to $\mu_{\text{hf--hf}}$
is the fact that owing to the low concentration of hard frictional grains, the probability of direct contacts of two hard grains is substantially smaller than that of a HF sphere and a hydrogel neighbor.

It is worth mentioning, that the contact model used in our study does not quantitatively reproduce the characteristic time of the unstable clogs obtained experimentally with low-frictional hydrogels. In fact, the Hertz-Mindlin contact model is more suitable for reproducing the behavior of hard grains, where clogs are typically very stable. However, both experiment and simulations are in good qualitative agreement, and allow a comprehensive analysis of the system from complementary points of view. This study demonstrates that the behavior of mixtures of grains with different frictional and elastic properties cannot be described by a simple interpolation of dynamic parameters, but that already the presence of a low percentage of one of the species can alter the dynamics substantially. In many practical situations, for instance in agriculture, but also in natural phenomena like mud or debris flow, the coexistence of  grains of very different sizes and mechanical properties is the rule, not the exception. It adds a new level of complexity that has been studied only scarcely in the past. Here, we have peeked into this complex field by restricting our study to monodisperse materials, in a quasi-2D geometry. We also restricted the discussion to small doping percentages. This is a considerable simplification of relevant practical cases, but it shines a light on the various phenomena expected in such systems.
Finally, further investigations should be carried out, addressing the significance of adding \rs{rigid} grains in the evolution of the force chains and contact network.

\section*{Conflicts of interest}

The authors declare no conflict of interest.

\section*{Acknowledgments}
This project has received funding from the European Union's Horizon 2020 research and innovation programme under the Marie Sk\l{}odowska-Curie
grant agreement {\sc CALIPER} No 812638. The Spanish MINECO (FIS2017-84631-P MINECO/AEI/FEDER, UE Projects) and the Hungarian NKFIH (grant No. OTKA K 116036) supported this work. The authors acknowledge discussions with J. Dijksman and J. van der Gucht.
The content of this paper reflects only the authors' view and the Union is not liable for any use that may be made of the information contained therein.\smallskip

\noindent\includegraphics[width=0.09\textwidth]{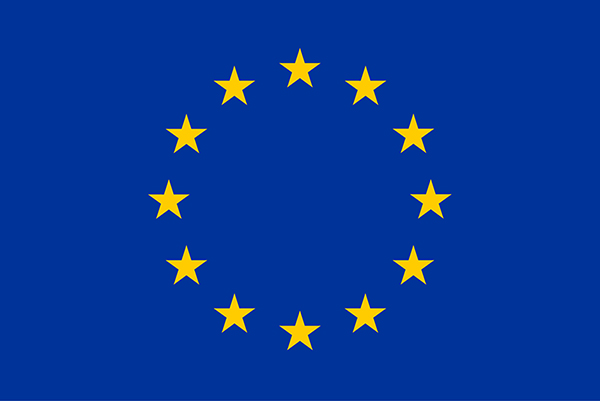}

\section*{Appendix A}

The statistics of avalanche size distributions for hard grain samples can be described
following the model of Thomas and Durian \cite{Thomas2013}. They assume that the
probability $p$ that a clog is formed when the $n$-th particle is
discharged is constant. Then, the probability that a clog stops
an avalanche of size $s$ is
\begin{equation}\label{eq:1}
    p_s= (1-p)^{s-1}\,p = p\exp\left(- \frac{s-1}{s_0}\right).
\end{equation}
(the first particle passes with probability one, otherwise there is no avalanche)
with $s_0=-1/\ln(1-p)$. The mean avalanche size $\langle s\rangle$ amounts to
\begin{equation}
  \langle s\rangle  = \sum_{s=1}^\infty s \, p_s = \frac{1}{p}  .
\end{equation}
This quantity is a characteristic parameter often used in the description of silo discharge statistics. For all practical situations (i.~e. $\langle s\rangle\gg 1$), both quantities
$\langle s\rangle$  and $s_0$ can be considered equal.
Experimentally, it is easier to evaluate the cumulative probability  $\Pi(S)$ of avalanches
having a size larger than $S$
\begin{equation}\label{eq:Cum1}
\Pi (S) = (1-p)^{S}.
\end{equation}
($S$ particles have passed the orifice with probability $(1-p)$ each).
The avalanche statistics is found
straightforwardly from a fit of the exponent in
$
   \Pi (S) = \exp\left(- \frac{S}{s_0}\right).
$ 
In a mixture of hard and elastic grains, one can assign different blocking probabilities $p_{\rm hf}$ and $p_{\rm e}\equiv p_{\rm hyd}$ to the respective components.

For soft grains, the blocking probability is a function of the container fill height \cite{Ashour2017b}. When the pressure at the orifice is high, $p_{\rm hyd}$ is small, and it increases with lowering pressure. The consequence is that the avalanche size distribution changes with fill height, the avalanches become shorter on average while the container empties.
Yet in any case, $p_{\rm hyd}$ is considerably lower than for hard grains of the same diameter.
Experiments with mixtures can provide the relative occurrences of hard and soft
grains in the blocking arches, which yield a quantitative measure of relative blocking probabilities.






\providecommand*{\mcitethebibliography}{\thebibliography}
\csname @ifundefined\endcsname{endmcitethebibliography}
{\let\endmcitethebibliography\endthebibliography}{}

\end{document}